\documentclass[12pt,preprint]{aastex}

\newcommand{\etal}{{et~al.\,}}
\newcommand{\mic}{$\mu$m$\,$}
\newcommand{\teff}{T$_{\rm eff}\,$}

\begin{document}
\sloppy
\slugcomment{Accepted to Ap.J.}

\title{Line Intensities and Molecular Opacities of the FeH $F^4\Delta_i-X^4\Delta_i$
Transition}

\author{M. Dulick\altaffilmark{1}, C. W. Bauschlicher, Jr.\altaffilmark{2}, Adam
Burrows\altaffilmark{3}, C. M. Sharp\altaffilmark{3}, R.S.
Ram\altaffilmark{4} and Peter Bernath\altaffilmark{4,5}}

\altaffiltext{1}{National Solar Observatory, 950 N. Cherry Ave., P.O. Box 26732,
Tucson, AZ \ 85726-6732 }
\altaffiltext{2}{NASA Ames Research Center, Mailstop 230-3, Moffett Field, CA \ 94035; charles.w.bauschlicher@nasa.gov}
\altaffiltext{3}{Department of Astronomy and Steward Observatory, The University of Arizona, Tucson, AZ \ 85721; burrows@jupiter.as.arizona.edu, csharp@as.arizona.edu}
\altaffiltext{4}{Department of Chemistry, The University of Arizona, Tucson, AZ
\ 85721; rram@u.arizona.edu}
\altaffiltext{5}{Department of Chemistry, University of Waterloo, Waterloo, Ontario, Canada  N2L 3G1; bernath@uwaterloo.ca}
\begin{abstract}

We calculate new line lists and opacities for the
$F^4\Delta_i-X^4\Delta_i$ transition of FeH. The 0-0 band of this
transition is responsible for the Wing-Ford band seen in M-type
stars, sunspots and brown dwarfs. The new Einstein A values for
each line are based on a high level ab initio calculation of the
electronic transition dipole moment. The necessary rotational line
strength factors (H\"onl-London factors) are derived for both the
Hund's case (a) and (b) coupling limits. A new set of
spectroscopic constants were derived from the existing FeH term
values for v=0, 1 and 2 levels of the $X$ and $F$ states. Using
these constants extrapolated term values were generated for v=3
and 4 and for $J$ values up to 50.5. The line lists (including
Einstein A values) for the 25 vibrational bands with v$\leq$4 were
generated using a merged list of experimental and extrapolated
term values. The FeH line lists were use to compute the molecular
opacities for a range of temperatures and pressures encountered in
L and M dwarf atmospheres. Good agreement was found between the
computed and observed spectral energy distribution of the L5 dwarf
2MASS-1507.

\end{abstract}

\keywords{infrared: stars --- stars: fundamental parameters --- stars: low mass, brown dwarfs, spectroscopy,
atmospheres, spectral synthesis}

\section{INTRODUCTION}

Wing and Ford (1969) were the first to detect a mysterious band
near 9910 \AA\, in late M dwarfs on the basis of low resolution
spectra. The Wing-Ford band was later found in S-stars \citep{ww2}
and in sunspots at higher spectral resolution by Wing, Cohen \&
Brault (1977). Nordh, Lindgren and Wing (1977) identified the
Wing-Ford band as the 0-0 band of a FeH electronic transition by
comparison with an unassigned laboratory spectrum of FeH that
showed a head at 9896 \AA\, degraded to longer wavelengths.
Laboratory spectra of the 1-0 band of this FeH transition show a
head at 8691 \AA\, (Carroll, McCormack \&\ O'Connor 1976) and this
band is also seen in sunspots (Wing, Cohen \&\ Brault 1977). Fawzy
et al. (1998) have detected the 2-0 and 2-1 bands in sunspots.

FeH bands can be observed well past 1 \mic with the 0-1 band of
the $F^4\Delta_i - X^4\Delta_i$ transition easily visible near
1.19 \mic in sunspots (Wallace \& Hinkle 2001a). A new electronic
transition of FeH has been identified near 1.583 \mic in late M
dwarfs, early L dwarfs (Leggett \etal 2001; Cushing \etal 2003),
and in sunspots (Wallace \& Hinkle 2001b). Although not assigned
yet, the 1.583 \mic band is probably the ${E^4\Pi} - {A^4\Pi}$
transition of FeH on the basis of a comparison with a laboratory
spectrum and the ab initio calculation of Langhoff and
Bauschlicher (1990). Wilson and Brown (2001) have tentatively
identified two spin components of the $A^4\Pi$ state about 1000
cm$^{-1}$ above the $X^4\Delta_{7/2}$ spin component by low
resolution dispersed laser-induced fluorescence.

On the astronomical front, there have also been recent advances in FeH
observations. The Wing-Ford band has been found in wide variety of sources,
including galaxies (e.g., Hardy \& Couture 1988). The most important application
of the ${F^4\Delta_i} - {X^4\Delta_i}$ transition is, however, in the spectroscopy of
L-type brown dwarfs. Indeed the presence of the metal hydrides (FeH and CrH) and
the absence of the metal oxides (TiO and VO) is the defining characteristic of
L-type brown dwarfs \citep{kir,kirb,kir20}. The main source of opacity
near 1 \mic is the ${A^6\Sigma^+} - {X^6\Sigma^+}$ transition of CrH and the
${F^4\Delta_i} - {X^4\Delta_i}$ transition of FeH.

Missing in all of the work on FeH is a measurement of the
oscillator strength of any of the FeH transitions. Without line
strengths, it is not possible to determine the column densities of
FeH from observations and the physical properties of the
atmospheres also depend on the FeH opacity. Schiavon, Barbuy and
Singh (1997) have provided some line strengths for the
${F^4\Delta_i} - {X^4\Delta_i}$ transition by arbitrarily choosing
an oscillator strength of 10$^{-3}$, at variance with the value
calculated by Langhoff and Bauschlicher (1990). In addition, they
seem to have made some computational errors, perhaps associated
with the degeneracy factors. We decided, therefore, to apply to
FeH the methods we used in our recent calculation of the line
strengths and opacity of the ${A^6\Sigma^+} - {X^6\Sigma^+}$
transition of CrH \citep{bau,bur}. We thus carried out a new ab
initio calculation of the transition dipole moment of the
${F^4\Delta_i} - {X^4\Delta_i}$ transition of FeH. In addition, we
have derived analytical expressions for the H\"{o}nl-London
rotational line strength factors for ${^4\Delta} - {^4\Delta}$
transitions in both the Hund's case (a) and Hund's case (b)
limits. The existing term values of Phillips \etal (1987) were
extrapolated to higher J and v based on a new determination of the
spectroscopic constants.

In the laboratory, early work in Ireland \citep{mcc} and Sweden
\citep{kly} provided line lists but no rotational assignments.
The Wing-Ford band was established as a ${^4\Delta} - {^4\Delta}$ transition by
rotational analysis of the 1-0 and 0-0 bands of FeD by Balfour, Lindgren \&
O'Connor (1983). The more heavily perturbed FeH bands finally yielded to
rotational analysis by Phillips \etal  in 1987, who assigned the 2-0 (7786
\AA), 1-0 (8692 \AA), 2-1 (9020 \AA), 0-0 (9896 \AA), 1-1 (10253 \AA), 0-1
(11939 \AA) and 1-2 bands (12389 \AA). These bands were recorded by Fourier transform
emission spectroscopy using a carbon tube furnace at 2200 - 2450$^\circ$C with 75
- 500 torr of H$_2$ and He. The Phillips \etal (1987) paper is still the
definitive work on the 1 \mic bands of FeH.

It was not clear in the early work that the lower state of the
Wing-Ford band was, in fact, the ground state of FeH. Both ab
initio calculations \citep{wsp} and photoelectron spectroscopy of
FeH$^-$ (Stevens, Feigerle \&\ Lineberger 1983) provided strong
evidence for a X$^4\Delta$ ground state. Langhoff and Bauschlicher
(1990) located all of the low-lying electronic states by ab initio
calculation and suggested the label F${^4\Delta_i} -
{X^4\Delta_i}$ for the 1-\mic band system. (The right subscript
$i$ stands for inverted and indicates the four spin components
with $\Omega$ = 7/2, 5/2, 3/2, 1/2 increase in energy as $\Omega$
decreases.) A recent calculation of the properties of the
low-lying electronic states has been carried out by Tanaka \etal
(2001).

More recent laboratory work on FeH includes the estimation of the
dissociation energy as 1.59 $\pm$ 0.08 eV (at 0 K) by Schultz and
Armentrout (1991), and the calculation of the first rotational
transition J = 9/2 $\leftarrow$ 7/2 $(v^{\prime\prime} = 0,
X^4\Delta_{7/2}$) at 1411.0927 GHz ($a-a$) and 1411.3579 GHz
($b-b$) on the basis of pure rotational laser magnetic resonance
spectroscopy in the far infrared region (Brown, Beaton \&\ Evenson
1993). The ground X$^4\Delta_i$ state has anomalously large
$\Lambda$-doubling, presumably due to the interaction with the
nearby A$^4\Pi$ state some 1200 cm$^{-1}$ above \citep{lan}.
Phillips \etal (1987) chose to follow the Brown \etal (1975)
recommendation of using the label $a$ for the lower energy level
of a $\Lambda$ doublet and $b$ for the upper level when the actual
$e/f$ parity label could not be ascertained. We will also adopt
this $a/b$ labelling scheme. J. Brown (private communication) has,
in fact, been able to determine the absolute parity in the ground
state and he finds that $e$ lies below $f$ for all four spin
components (i.e., $a=e$ and $b=f$).

The mid-infrared laser magnetic resonance spectra of FeH were recorded by Towle
et al. (1993) and they improved on a few of the predicted line positions
\citep{phi} of the 1-0 vibration-rotation band of the
X$^4\Delta_{7/2}$ spin component. Phillips \etal (1987) provided a list of term values for v=0, 1
and 2 of the F$^4\Delta_i$ and X$^4\Delta_i$ states determined from an \AA slund
term value analysis
of the line positions. These term values were confirmed by Towle \etal (1993)
to have an accuracy of $\pm$0.02 cm$^{-1}$.

The infrared spectrum of FeH (along with FeH$_2$ and FeH$_3$) in an argon matrix was
recorded by Chertihin and Andrews (1995). Evidently, the earlier matrix
measurements by Dendramis, Van Zee and Weltner (1979) were of FeH$_2$ and
not of FeH.

In addition to the $F^4\Delta_i - X^4\Delta_i$ transition, FeH has visible electronic
transitions in the blue (4920 \AA) and green (5320 \AA) (Carroll \&\ McCormack
1972; Carroll,
McCormack \&\ O'Connor 1976) that probably also occur in sunspots.
However, these FeH lines are heavily blended in the sunspot spectrum, in
contrast to the Wing-Ford band. Brown and co-workers (for example, Wilson \&
Brown 2001; 1999; Hullah, Barrow \& Brown 1999) have devoted much time to the
analysis of the visible spectra of FeH using high resolution dye lasers and a
much cooler (450 K) source of molecules. Some of the visible bands (e.g.,
Hullah, Barrow \& Brown, 1999) connect to the ground state and improve and
extend a few of the term values of Phillips \etal (1987).

\section{AB INITIO CALCULATION OF BAND STRENGTHS}
\nobreak
The experimental measurement of the radiative lifetime of the Wing-Ford band is
difficult mainly because 1 $\mu$m is not a convenient wavelength and FeH is
difficult to make in the laboratory. In the absence
of experimental work, we chose to calculate the transition dipole moment
function from the wavefunctions obtained by ab initio solution of the electronic
Schr\"{o}dinger equation.

The orbitals are determined using a state-averaged complete-active-space self-consistent-field
(SA-CASSCF) approach that includes symmetry and equivalence restrictions.  In $C_{2v}$ symmetry,
the active space contains 6 $a_1$ orbitals and one orbital each of $b_1$, $b_2$, and $a_2$
symmetry.  This active space includes the Fe $3d$, $4s$, and $4p\sigma$ orbitals, the H $1s$ orbital,
and one extra $\sigma$ orbital, which is added on the basis of preliminary calculations.
The three lowest $^4\Delta$ states are included in the SA-CASSCF approach.  The third state is
added to smooth the potentials and transition moment at larger r values.  While the addition
of the third state improves the description at large r values, it makes only a very small difference
in the final potential at shorter r values.  More extensive
correlation is included using the multireference configuration interaction approach, MRCI.
These MRCI calculations are performed in $C_{2v}$ symmetry for the four lowest states,
which include the three $^4\Delta$ states and a low-lying $^4\Sigma^-$ that is in the
same $C_{2v}$ representation.

The MRCI calculations use all CASSCF configurations as reference configurations and
internal contraction~\citep{ic1,ic2} (IC) is used to limit the size of the calculations.
The importance of higher excitations is estimated using the multireference analog of the
Davidson correction, which is denoted +Q.  The inclusion of Fe $3s$ and $3p$ correlation did not significantly
improve the potentials and, therefore, is not included in the final calculation that
correlates the Fe 3d and 4s and H 1s electrons.  Scalar relativistic effects are included
using the one-electron Douglas Kroll (DK) approach\ \citep{hess}.

The Fe basis set can be described as (20s15p10d6f4g)/[7s7p5d3f2g]. The primitive set is derived
from that of Partridge \citep{hrp}.  The s, p, and d primitives are contracted based on an DK-SCF
calculation of the $^5D$ state of Fe atom.  The inner 16 s primitives are contracted to 3 functions
and the outermost 4 are uncontracted, the inner 10 p primitives are contracted to 2 functions and
the outermost 5 are uncontracted, and the inner 6 d primitives are contracted to 1 function and the outermost
4 are uncontracted.  The f and g sets are taken from our averaged atomic natural
orbital set \citep{aano}.
For hydrogen, the augmented-correlation-consistent polarized-valence triple zeta
(aug-cc-pVTZ) set \citep{cc1,cc2}
is used, but the contraction coefficients are taken from a DK-SCF calculation.
The CASSCF/IC-MRCI calculations are performed using Molpro \citep{molpro}.

The computed IC-MRCI+Q spectroscopic constants are tabulated in
Table 1.  The IC-MRCI+Q results are in reasonable agreement with
present experimental values (derived below). Since it is difficult
to perform a more accurate calculation and the Rydberg-Klein-Rees
(RKR) potentials developed from the experimental constants have
problems with the inner wall at relatively low energies, we
combine the computed results with experiment to develop our final
potentials.  We shift the computed potentials so that the computed
$B_0$ values agree with experiment: the ground state is shifted to
longer $r$ by 0.0218~\AA\ and the excited state to shorter $r$ by
0.0270~\AA.  The upper state is shifted up in energy so that the
computed $T_0$ value agrees with the experimental (7/2-7/2) 0-0
band origin.  We then generate RKR potentials using the computed
$B_0$, $B_1$, and $B_2$ values from the shifted potentials and the
experimental $\omega_e$ and $\omega_e x_e$ values (averaged over
the two lambda doublet components).  Using these RKR potentials
and the IC-MRCI transition moment, we evaluate the Franck-Condon
factors and Einstein A values (see Table 2).

\newcommand{\bfm}[1]{\mbox{\bf{#1}}}
\newcommand{\wtj}[6]{\left(\begin{array}{ccc}
#1 & #2 & #3\\[5pt]
#4 & #5 & #6
\end{array}\right)}
\newcommand{\wsj}[6]{\left\{\begin{array}{ccc}
#1 & #2 & #3\\[5pt]
#4 & #5 & #6
\end{array}\right\}}
\newcommand{\qn}{\Lambda S \Sigma J M \Omega}
\newcommand{\qnp}{\Lambda^\prime S^\prime \Sigma^\prime J^\prime M^\prime
\Omega^\prime}

\section{CALCULATING H\"{O}NL-LONDON FACTORS FOR A $^4\Delta$ --
$^4\Delta$ TRANSITION}

Rotational line strength factors are needed to compute the individual line
intensities. Because none is available in the literature for a $^4\Delta -
^4\Delta$ transition, we have derived the H\"{o}nl-London factors needed.

The elements of an electric-dipole transition matrix \bfm{T} for a
Hund's case (a) $\qnp \rightarrow \qn$ rotational transition in
the absence of external fields can be expressed in terms of
compact spherical-tensor notation as

\begin{eqnarray}
&<&\qnp\:|\:\mbox{\bf T}\:|\:\qn> \nonumber\\
\nonumber\\
& = & (-1)^{J^\prime - \Omega^\prime}\:
\wtj{J^\prime}{1}{J}{-\Omega^\prime} {q}{\Omega}\:
<\Lambda^\prime\:||\:{\mbox{\boldmath$\mu$}}\:||\:\Lambda>
\delta(S,S^\prime)\:\delta(\Sigma,\Sigma^\prime)\:\delta(M,M^\prime).
\end{eqnarray}

Here {\mbox{\boldmath$\mu$}} = -e\bfm{r} is the electric-dipole
operator and for case (a) coupling, $\Omega = \Lambda + \Sigma$.
The presence of the delta functions for $S$ and $\Sigma$ occur
because {\mbox{\boldmath$\mu$}} does not depend on electron-spin
coordinates. The delta functions and the conditions for a nonzero
Wigner 3-$j$ symbol establish the well-known case (a) selection
rules for allowed transitions:

\begin{eqnarray*}
 &\Delta S = \Delta\Sigma = 0,&\\[5 pt]
 &\Delta\Omega = -q = 0,\:\pm 1,&\\[5pt]
 &\Delta J = -1\:\mbox{(P-branch)},\:0\:\mbox{(Q-branch)},\:\mbox{and}\:+1
\:\mbox{(R-branch)}&
\end{eqnarray*}

The factor $<\Lambda^\prime\:||\:{\mbox{\boldmath$\mu$}}\:||\:\Lambda>$ is a constant
for a given $\Lambda^\prime \rightarrow \Lambda$ transition. The only
real importance of this factor here is that it establishes the selection
rule, $\Delta\Lambda = 0, \pm 1$, due to the fact that {\mbox{\boldmath$\mu$}} is a
rank-one spherical tensor. Furthermore, $\Delta\Omega = \Delta\Lambda
+ \Delta\Sigma$, and since $\Delta\Sigma = 0$, the selection rule for
spin-component transitions is reduced simply to $\Delta\Omega = \Delta\Lambda$.

The remaining property to fully characterize case (a) rotational
levels is parity. In recent decades, the $e/f$ parity scheme has
become the conventionally accepted method by molecular
spectroscopists for parity labelling the case (a) rotational
levels. The transition matrix in the $e/f$ parity basis is
obtained by the transformation,

\begin{eqnarray}
\bfm{U}^\dagger\:\bfm{T}\:\bfm{U} & &
\end{eqnarray}

\noindent
where $\bfm{U}$ is the Wang matrix,

\begin{eqnarray}
U_{ij} &=& \left\{
\begin{array}{l@{\hspace{20pt}}l}
 -1 / \sqrt{2} & j = i < n/2\\[10pt]
 +1 / \sqrt{2} & j = i > n/2\\[10pt]
 +1            & j = i = n/2\\[10pt]
 +1 /\sqrt{2}  & j = n - i + 1\\[10pt]
  0            & \mbox{otherwise}
\end{array}\right.
\end{eqnarray}

\noindent
with $n$ being the total number of $\Lambda S \Sigma$ states.
For P and R branch lines the resultant matrix is partitioned into
two nonzero diagonal blocks of equal dimension, where one block corresponds
to $e$ $\rightarrow$ $e$ transitions and the other to $f$ $\rightarrow$ $f$.
In contrast, the transition matrix for the Q branch lines in the parity
basis is comprised of two nonzero off-diagonal blocks, the $e$ $\rightarrow$ $f$
and the $f$ $\rightarrow$ $e$ transitions.

Results for the $^4\Delta$ -- $^4\Delta$ transition ($\Omega^\prime$
= $\Omega$ = 1/2, 3/2, 5/2, 7/2) are displayed in Table 3. Apart from the
constant factor $<\!\Delta\:||\:{\mbox{\boldmath$\mu$}}\:||\:\Delta\!>$, the H\"onl-London
factors, designated by $S(J)$, are the squares of the matrix elements of
$\bfm{U}^\dagger\:\bfm{T}\:\bfm{U}$.

Because Hund's case (a) and case (b) are both well-defined coupling schemes, that is,
the coupling does not depend on molecular parameters, it then becomes a
matter of just simply constructing a transformation matrix \bfm{U}
in closed analytical form where $\bfm{U}^\dagger\:\bfm{T}\:\bfm{U}$ converts
\bfm{T} in the case (a) nonparity basis over to case (b), or vice versa. The matrix
elements for such a transformation are

\begin{eqnarray}
& < & \qn|\Lambda N\Lambda SJM> = <\Lambda N\Lambda SJM|\qn\: > \nonumber\\
\nonumber\\
& = & \: (-1)^
{S-N+\Omega}\:\sqrt{2N+1}\:\wtj{J}{S}{N}{-\Omega}{\Sigma}{\Lambda}.
\end{eqnarray}

The case (b) H\"onl-London factors in Table 4 were derived using this
approach.

For case (b) coupling, $\Sigma$ and $\Omega$ are no longer defined
whereas $\Lambda$ is a well-defined quantum number for both
coupling cases. Secondly, rotational energies for case (a) are
computed using the Hamiltonian operator $B(r)\:\bfm{R}^2$, where
$\bfm{R} = \bfm{J} - \bfm{L} - \bfm{S}$. For case (b) the
Hamiltonian operator is $B(r)\: \bfm{N}^2$, where $\bfm{N} =
\bfm{R} + \bfm{L} = \bfm{J} - \bfm{S}$. Thus, for a given $J$
value and discounting parity, there are four rotational levels
with $N$ values, $J-3/2$, $J-1/2$, $J+1/2$, and $J+3/2$ for $S =
3/2$, which equals the number of spin-components with $\Omega$
values, 1/2, 3/2, 5/2, and 7/2 for case (a) coupling. Also, the
$e/f$ parity labelling is not convenient for the case (b)
rotational levels since the basis functions $|\Lambda N\Lambda
SJM>$ are already eigenfunctions of the parity operator; they are
labelled simply as $+$ or $-$ to correspond to the $\pm\Lambda$
degeneracy. For case (b) rotational transitions, the parity
selection rules are $+ \rightarrow +$ and $- \rightarrow -$,
regardless of whether the branch is P, Q, or R. This + or $-$
orbital parity is written as a right superscript to the term
symbol (e.g., $^1\Sigma^+$ or $^4\Delta^+$) and is not to be
confused with total parity (Bernath 1995).

These finer points between case (a) and (b) help to explain the
notation differences in Tables 3 and 4. The $\Omega$ and parity
labels along with the $\Delta J$ P, Q, and R branch designations
in Table 3 are sufficient to uniquely label all the allowed case
(a) rotational transitions. The quantum number $N$ in case (b)
serves as a label analogous to $\Omega$ in case (a). In addition
to the strong $\Delta N = \Delta J$ transitions, that form the
so-called main branches (analogous to the case (a) $\Delta\Omega =
\Delta\Lambda$ branches), there also exist weaker $\Delta N \neq
\Delta J$ transitions, which account for the better than two-fold
increase in the number of branches. To distinguish $\Delta N =
\Delta J$ from $\Delta N \neq \Delta J$ branches, the traditional
convention is to augment the $\Delta J$ P, Q, or R branch
designations with $\Delta N$ P, Q, or R left superscript
designations when $\Delta N \neq \Delta J$. In Table 4 this
convention is not strictly followed; to alleviate any confusion,
the redundant $\Delta N$ superscripts are retained for the main
branches as well. Note that in order to avoid errors in the
H\"onl-London factors in Tables 3 and 4, they were generated with
the computer algebra program Maple.

For the $X$ and $F$ $^4\Delta$ states of FeH, the coupling is intermediate
between case (a) and case (b) for increasing $J$. To further complicate
matters, the rotational levels for both of these states are perturbed by
unknown electronic states. The former problem of intermediacy can be treated
by constructing the energy Hamiltonian $\bfm{H}$ in the case (a) basis as
shown in Table 5. The eigenvectors $\bfm{U}$ from the diagonalization
$\bfm{U}^\dagger\:\bfm{H}\:\bfm{U}$ for both the lower and upper states
$l$ and $u$ are then used to correct the H\"{o}nl-London factors for
intermediate coupling by the process $\bfm{U}^\dagger_l\:\bfm{T}\:\bfm{U}_u$.

The Phillips \etal (1987) tables of experimental term values for
the $X$ and $F$ $v$ = 0, 1, and 2 levels were used along with the
eigenvalues of $\bfm{H}$ in least-squares fits to determine the
molecular constants, $T_v$, $B_v$, $D_v$, $A_v$, and $\lambda_v$,
separately for each parity $a$ and $b$. (The distinction between
$e/f$ and the Phillips \etal $a/b$ labelling is irrelevant as far
as the analysis presented here is concerned.) Due to the
perturbations, these fits gave rather disappointing results,
yielding standard deviations on the order of 2 -- 4 cm$^{-1}$ for
the $X$ levels and 1 -- 3 cm$^{-1}$ for the $F$ levels. To
minimize the effects of these perturbations as much as possible,
an alternate fitting procedure was tried using the average of the
experimental term values for each J value for the four spin
components. The results from these fits showed remarkable
improvement with roughly a factor of 10 reduction in the standard
deviations, 0.1 -- 0.7 cm$^{-1}$ for the $X$ levels and 0.2 -- 0.4
cm$^{-1}$ for the $F$ levels. Nevertheless, these standard
deviations are still a factor of 10 higher than the estimated
experimental uncertainties of better than 0.05 cm$^{-1}$,
indicating the extent of residual effects due to the perturbations
that averaging was unable to remove. Fitting the term values in
this manner cannot determine the spin-orbit and spin-spin
constants. These $A$'s and $\lambda$'s instead were determined in
separate fits utilizing the spin-splitting data in Table 6C of
Phillips \etal (1987). To within the quoted experimental error of
0.5 cm$^{-1}$, the spin-splittings for all the observed
vibrational levels can be reproduced by a single $A$ and $\lambda$
value for each electronic state. The molecular constants in Table
7 represent a marginal improvement over those reported by Phillips
\etal, who derived their constants by fitting the rotational line
positions to a simple polynomial expression. Our new constants,
however, extrapolate somewhat better to higher $J$ and $v$ values.

The specified range in temperature for the opacity calculations
required extending the spectrum to include rotational lines up to
$J^{\prime\prime}$ = $J^\prime$ = 50.5 and vibrational transitions
up to $v^{\prime\prime}$ = $v^\prime$ = 4. The terms values used
in generating the spectrum for the low-resolution opacity
calculations are listed in Table 8. The entries are a combination
of experimental term values taken from the tables of Phillips
\etal and calculated values for the higher $v$ and $J$ levels. The
missing entries in their tables for the lower $J$ levels were
calculated by simple polynomial extrapolation. Even in spite of
the perturbations, the observed term values are surprisingly
smooth functions of $J$. This smoothness was maintained in
extending the term values to higher $J$ by averaging the observed
to calculated term value ratios in the region of overlap, and then
multiplying the calculated term values outside the region by this
average ratio.

In light of the perturbations, there are obvious concerns
regarding the reliability of the H\"{o}nl-London factors for
intermediate coupling (Tables 3 or 4) using either the case (a)
Hamiltonian model in Table 5 or case (b) model in Table 6. To
address this issue we did a comparison of the H\"onl-London
factors with the measured line intensities from a McMath-Pierce
FTS archived emission spectrum recorded on May 23, 1984 at a
furnace temperature of 2350$^\circ$C and a background pressure of
70 Torr. This spectrum was then ratioed to the archived companion
Optronics reference lamp spectrum to remove the variation of
detector response with wavenumber from the background signal. The
bands specifically targeted for this comparison were the $\Delta
v$ = 0 and $-1$ bands involving the main $\Delta\Omega=0$ branches
vs. the $\Delta\Omega=\pm 1$ satellite branches. Rotational line
intensities in these bands were measured by fitting the observed
rotational lineshapes to a Voigt lineshape function preceded by
subtraction of the background signal level in the vicinity of line
center to determine absolute peak intensities. Measured rotational
line intensities in the satellite branches were for the most part
a factor of 6 times weaker than the corresponding ones in the main
branches that shared common upper levels. The calculated line
strength factors based on the Table 5 or 6 Hamiltonian model
predict on the other hand that the ratio of these satellite to
main branch lines should be weaker by a factor more like 70 for
the low-$J$ lines ($J^{\prime\prime} \sim 3.5$) and 50 for the
high-$J$ lines ($J^{\prime\prime} \sim 50.5$). Obviously the $F$
and $X$ states are not pure $^4\Delta$ states.

\section{GENERATION OF LINE LISTS}

The line positions for the ${F^4\Delta_i}-{X^4\Delta_i}$ transition for v=0-4
and J $\leq$ 50.5 were generated from term values in Table 8. All possible lines were
computed consistent with the electric dipole selection rules for J ($\Delta J=0,
\pm1$) and parity ($a-a$, $b-b$ for R, P branches and $a-b$, $b-a$ for Q branches). For
Hund's case (a) coupling there are only two strong branches (P and R) and a Q
branch whose intensity decreases rapidly with J (Table 3) for each spin component
(12 branches in total). For Hund's case (b), numerous satellite branches appear
and there are 28 branches in all.

The Einstein A values for each line were computed using the
formula \linebreak $A=A_{v^{\prime}v^{\prime \prime}}$ HLF
/($2J^{\prime} + 1$), in which the Einstein A for the
$v^{\prime}-v^{\prime \prime}$ band is taken from Table 2 and the
H\"{o}nl-London factor (HLF) is computed as discussed above.
Because the $F-X$ transition of FeH is intermediate between Hund's
case (a) and (b), the HLF factors were calculated from the Hund's
case (a) H\"{o}nl-London factors and the eigenvectors used to
diagonalize the Hamiltonian matrix. Note that at high $J$ these
calculated factors are nearly identical with those given by the
Hund's case (b) formula in Table 4. The individual Einstein $A$
values for each rovibronic line can be converted to dimensionless
$gf$ values with the formula:
\begin{equation}
gf = (2J^{\prime\prime}+1)f_{abs} = \frac{\epsilon_{0} m_{e}
c^3}{2 \pi {e}^2 {\nu}^2} A(2J^{\prime}+1) \label{eqgf}
\end{equation}
using SI units.

The line positions and line intensities were collected into 25
files, one for each of the bands involving $v^{\prime}$ and
$v^{\prime \prime}$ 0-4 and for P, Q, and R branches. These files
can be obtained from the web site http://bernath.uwaterloo.ca/FeH.

\section{CALCULATION OF FeH OPACITIES}
\label{opac}

As with the molecule CrH (Burrows et al. 2002), the integrated line
strengths were computed from input Einstein $A$ coefficients obtained
from the calculated line lists. Since the $A$ coefficients already include all the
details of the line strength, i.e. the strength of the associated band
and its Franck-Condon factor, as well as the individual H\"onl-London
factor, these factors do not need to be considered separately.
As in Burrows et al. (2002), the integrated line strength in
cm$^2$ s$^{-1}$ molecule$^{-1}$ is
calculated from
\begin{equation}
S = \frac{1}{8\pi\bar{\nu}^2} A(2J^{\prime}+1)
\exp(-E^{\prime\prime}hc/kT) (1-\exp(-hc\bar{\nu}/kT))/Q\, ,
\label{eqS}
\end{equation}
where $J^{\prime}$ is the upper rotational quantum number,
$\bar{\nu}$ is the transition wavenumber in cm$^{-1}$,
$E^{\prime\prime}$ is the excitation energy of the lower state in
cm$^{-1}$, and the $Q$ is the internal partition function. The
stimulated emission factor is also included in eq. (5). Note that
there is a small error in the Burrows et al. (2002) equation for
$S$ and the correct version has $2J^{\prime}+1$ not
$2J^{\prime\prime}+1$. The reason for this change is that an extra
factor of $(2J^{\prime}+1)/(2J^{\prime\prime}+1)$ is needed to
convert the emission quantity $A$ to the absorption quantity $S$.

The monochromatic cross section per molecule is obtain by
multiplying the value from eq. \ref{eqS} by a truncated Lorentzian
profile.  A simple prescription provided by R. Freedman (2000,
private communication) is used to calculate the J-dependent
full-width-half-maximum in cm$^{-1}$ of a line given by:
$$\Delta\bar{\nu} = [W_a - W_b min(J^{\prime\prime},30)] P_{atm}\, ,$$
where $W_a$ and $W_b$ are line broadening coefficients, with the
values 0.15 and 0.002, respectively, $J^{\prime\prime}$ is the
rotational quantum number of the lower state, and $P_{atm}$ is the
total gas pressure in atmospheres.  This simple prescription
causes the lines to decrease in width up to $J''=30$, and
thereafter remain at a constant width.

The line is profiled using a Lorentzian function on a spectral
grid with a 1 cm$^{-1}$ grid spacing.  In order to save computer
time and to simulate the rapid drop off in intensity in the far
wings, the profile is truncated beyond $\Delta\bar{\nu}_{trunc}$
cm$^{-1}$ from the line center, where $\Delta\bar{\nu}_{trunc}$ is
the smaller of $25 P_{atm}$ and 100, so a profile is never
computed more than 100 cm$^{-1}$ on either side of the line
center. However, in order to conserve the total line strength, a
normalizing factor is applied to the profile.

At very low pressures, the lines may be so narrow that they may
``fall between" the grid points and be undersampled or missed
completely.  To ensure that the lines are properly represented,
the line centers are moved to the nearest grid point, regardless
of the width, then for very narrow lines where only the grid point
at the line center is represented, the line is artificially
broadened to include the two neighboring grid points and
normalized so that the area covered corresponds to the total line
strength.

The internal partition function ($Q$) is calculated from all
electronic molecular states for which experimental and theoretical information
is available, and which make a significant contribution for the temperatures of
relevance. This includes the ground ($X^4\Delta$) and excited
($F^4\Delta$) states that produce the band systems being computed,
as well as 8 intermediate excited states that lie below the $F$
state. Of these 10 states, we have estimated the excitation energies of the
spin-split sub-states for the first 5. The energy levels and spectroscopic
constants used in the calculation of Q are provided in Table 9.

Associated with each of the 10 electronic states are the vibrational
and rotational constants $w_e$, $w_ex_e$, $B_e$ and $D_e$,
where it is assumed that for all spin-split sub-states
belonging to an electronic state, the constants are the same.
Normally for partition functions this is a good approximation. The
individual partition function for the electronic state $i$ is given by
the product $Q_{e_i} \times Q_{v_i} \times Q_{r_i}$, which are, respectively, the
electronic, vibrational, and rotational partition functions. $Q_{e_i}$
is a small integer depending on electron spin and orbital angular
momentum, and $Q_{v_i}$ and $Q_{r_i}$ are calculated using asymptotic
formulae from Kassel (1933a,b). The total internal partition function
is calculated by summing up the individual electronic states weighed
by the Boltzmann factor using

\begin{equation}
Q = \sum_{i=1}^n Q_{e_i} Q_{v_i} Q_{r_i} \exp(-T_ihc/kT)
\label{eqQ}
\end{equation}
where $T_i$ is the excitation energy of state $i$, and is zero for
the ground state ($i=1$). The computed partition function from
1000 K to 3500 K in 100 K increments is provided in Table 9. As a
test we computed the function by direct summation of the $X$ and
$F$ state energy levels used to generate the line list. The
partition function was also computed using the analytical formulas
(cited above) using all of the low-lying electronic states (Table
9). These low-lying electronic states make an important
contribution to $Q$ (Table 9) at all relevant temperatures and
should not be neglected.

In order to calculate the abundance of FeH at a particular
temperature and pressure in model calculations assuming chemical
equilibrium, the free energy of FeH is calculated at the
temperature in question, then the total free energy of the system
is minimized (Sharp and Huebner 1990). However, this requires
up-to-date thermodynamic data. Accordingly, these were calculated
from the partition function of FeH using eq. (\ref{eqQ}) with our
new data, the partition functions of the dissociated atoms Fe and
H, and the dissociation energy of FeH (taken as 1.598 eV). It was
found that the partition function calculated with our new data was
about a factor of two larger than the old value.

Along with the most abundant isotope ($^{56}$Fe, which makes up
91.7\% of iron), $^{54}$Fe, $^{57}$Fe, and $^{58}$Fe are included,
and make up 5.8\%, 2.2\%, and 0.28\%, respectively. The
corresponding vibrational, rotational, and vibration-rotation
coupling constants for any isotopically substituted molecule of
FeH can be obtained from the ratio of the reduced mass of that
molecule and the reduced mass of $^{56}$Fe$^{1}$H. These constants
depend on various 1/2-integer powers of the ratio of reduced
masses, as given by Herzberg (1950). From the vibrational and
rotational quantum numbers of the upper and lower states for each
transition, the isotopic shifts of the lines are computed. The
strength of each line is calculated by applying the fraction of
the isotope of iron to equation (\ref{eqS}). To calculate the
cross section as a function of frequency, this is then multiplied
by the abundance of FeH and the profile.

Due to the isotopic shifts of the energy levels, the partition
function and the Boltzmann factor of the lower states will be affected,
which will have a small effect on the line strengths. Likewise, there
will be a small effect due to a changed oscillator strength caused by
a shift in the frequency of the transition, and slightly changed
wavefunctions will alter the Franck-Condon factors. All these effects
are very small compared with the shifting of the lines, and so are ignored.
The shifted lines may be important as they could fill in
gaps in the $^{56}$Fe$^{1}$H opacity spectrum.

The whole process is repeated for each of the 25 bands of FeH considered
for all possible combinations of the upper and lower vibrational quantum
numbers taking values between 0 and 4, and for each of the 4 isotopic
versions; thus, 100 bands are calculated.

\section{REPRESENTATIVE FeH OPACITY PLOTS AND A THEORETICAL L DWARF SPECTRUM}
\label{opacplots}

\begin{figure}
 \epsscale{1.00}
\vspace*{-1.0in} \plotone{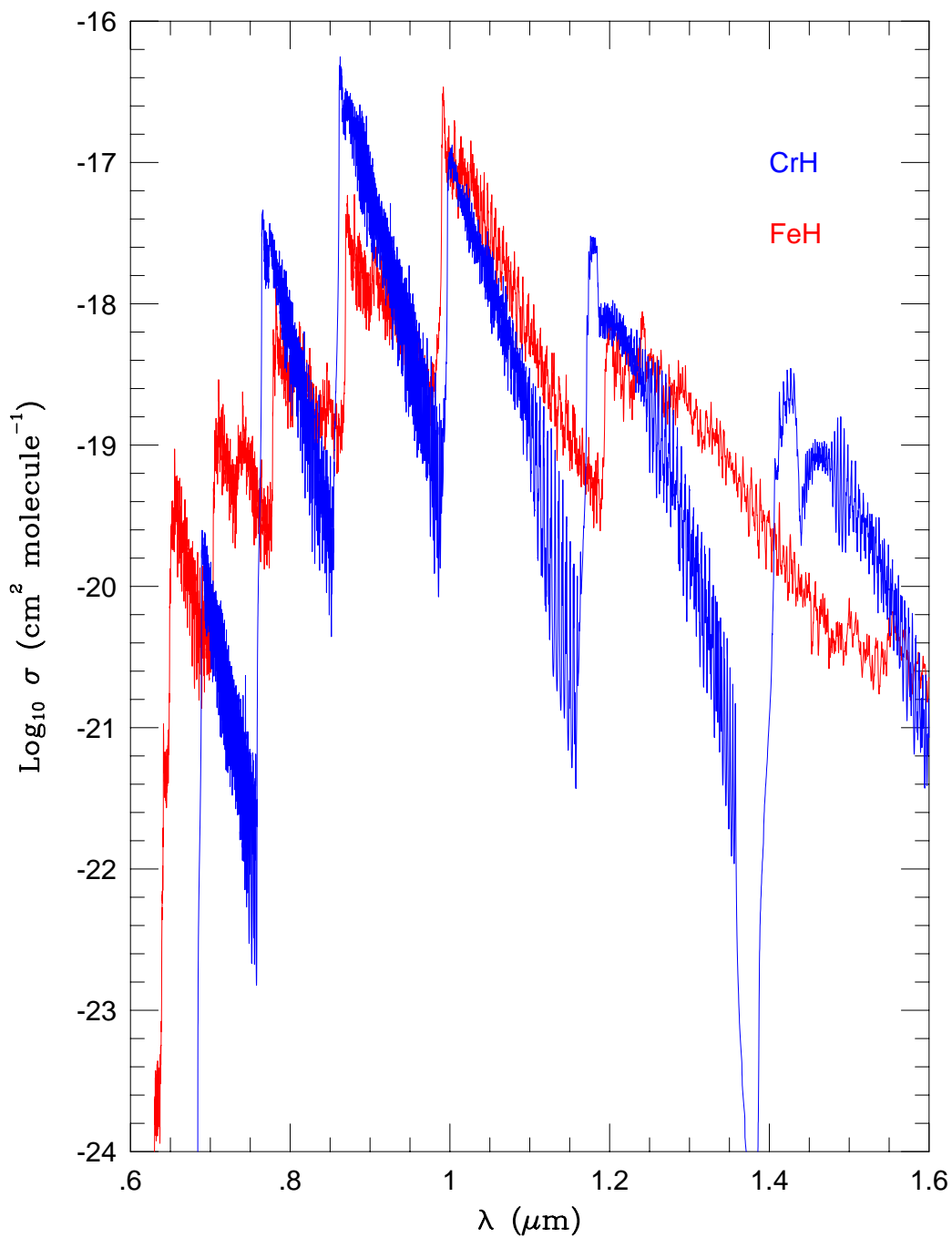} \vspace*{-1.0in} \caption{The
logarithm (base ten) of the absorption cross sections for FeH
(red) and CrH (blue), the former as derived in this paper, the
latter taken from Burrows et al. (2002).  The temperature and
pressure are 1800 Kelvin and 30 atmospheres, respectively.  Most
(but not all) of the important bands for both molecules are to be
found at wavelengths between 0.65 \mic and 1.6 \mic and these are
what are plotted here.  Notice that FeH and CrH frequently
contribute opacity in similar wavelength regions.  This is
particularly relevant around $\sim$0.86 \mic, $\sim$1.0 \mic, and
in the $J$ photometric band from $\sim$1.2 to $\sim$1.3 \mic.
\label{fig:1}}
\end{figure}

\begin{figure}
 \epsscale{1.00}
\vspace*{-1.0in} \plotone{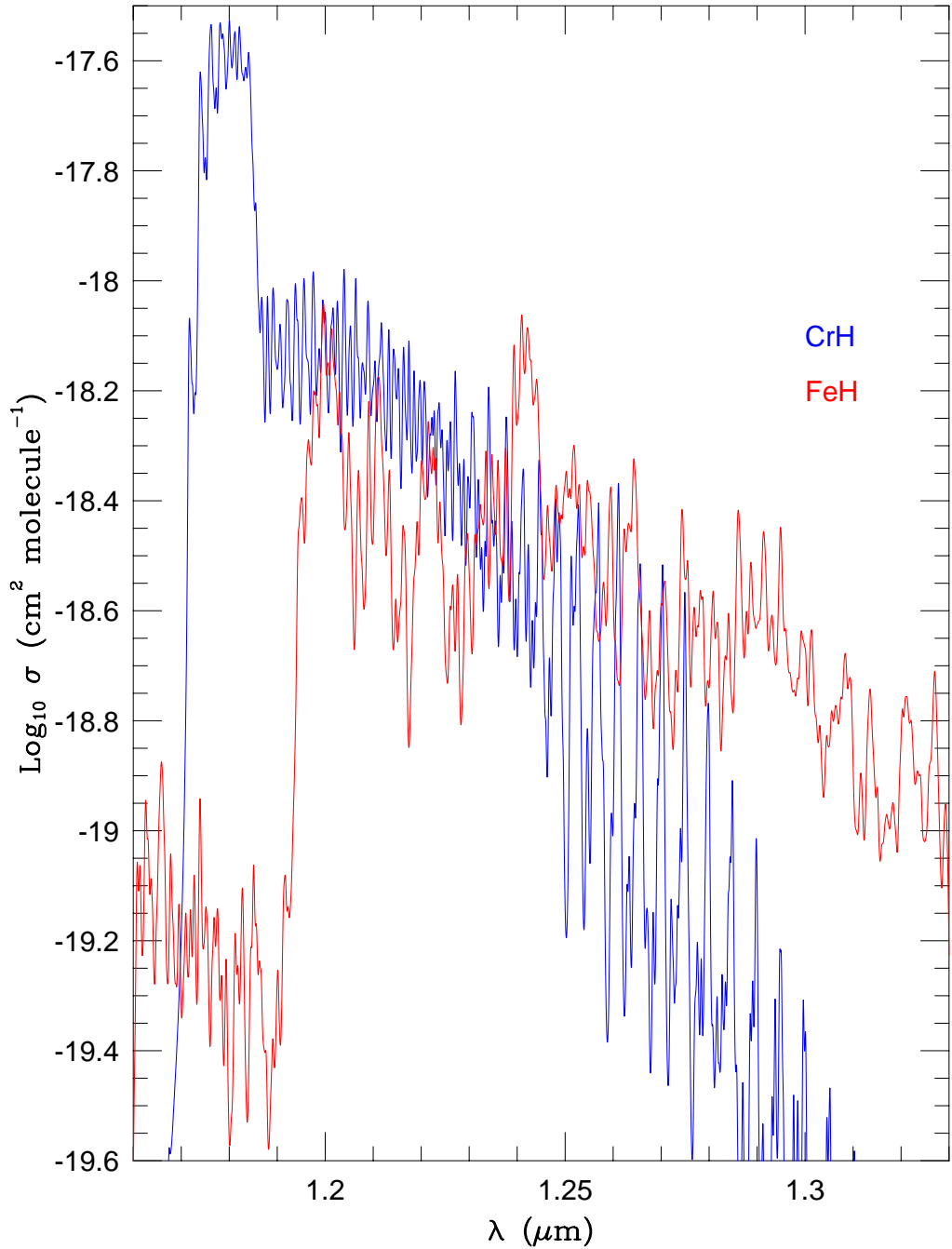} \vspace*{-1.0in} \caption{The
logarithm (base ten) of the absorption cross sections of both FeH
(red) and CrH (blue) in the $J$ band from 1.16 \mic to 1.33 \mic.
As this plot demonstrates, depending upon the relative abundances,
both FeH and CrH bands can contribute opacity and lines in the $J$
band.  A spectral resolution near 0.22 cm$^{-1}$ was employed.
\label{fig:2}}
\end{figure}

Using the procedure outlined in \S\ref{opac}, we have calculated tables of FeH
opacities for a broad range of the temperatures and pressures encountered in L
and T dwarf atmospheres.  These opacities have also been incorporated, along
with the new FeH abundances, into a spectral synthesis code (Burrows \etal 2002)
used to derive theoretical spectra and colors for cool, often substellar, objects with molecular
atmospheres.  A representative opacity spectrum at a temperature of 1800 K and a pressure
of 30 atmospheres for wavelengths from $\sim$0.65 \mic
to $\sim$1.6 \mic for both FeH and CrH (Burrows \etal 2002)
is provided in Fig. \ref{fig:1}.  The new FeH opacities are $\sim$100 times
larger than those found in Schiavon, Barbuy, and Singh (1997).  The CrH opacities are presented along with the
new FeH opacities to demonstrate the similarities and differences between the
opacities for the two molecules.  The Wing-Ford band of FeH near $\sim$0.99 \mic
is accompanied by a similar band for CrH. In addition, the prominent FeH
1-0 band feature near 0.87 \mic is accompanied by a CrH feature near 0.86 \mic. Importantly,
the $J$ photometric band between 1.19 \mic and 1.30 \mic boasts lines from both moleclues,
though previously only FeH features have been identified.
Figure \ref{fig:2} focuses on the $J$ band and has a resolution near 0.22 cm$^{-1}$.
Clearly both molecules must be considered when analyzing high-resolution
spectra in this band.

\begin{figure}
 \epsscale{1.00}
\vspace*{-1.0in} \plotone{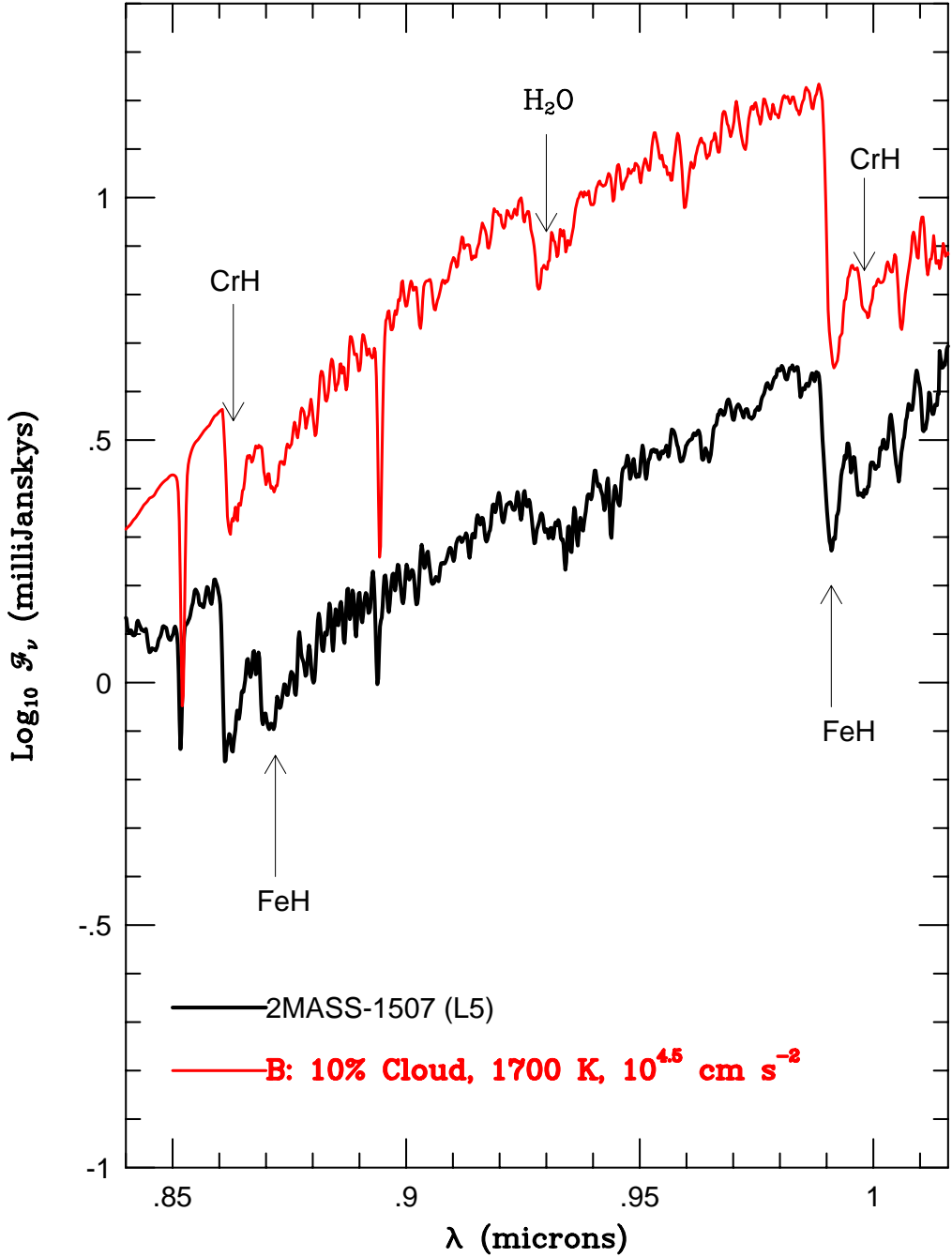} \vspace*{-1.0in} \caption{The
log (base ten) of the absolute flux density (${\cal F}_\nu$) in
milliJanskys versus wavelength ($\lambda$) in microns from 0.84
\mic to 1.02 \mic\ for a self-consistent theoretical
solar-metallicity model of the L5 dwarf 2MASS-1507.  The absolute
position of the theoretical spectrum, dependent as it is on the
object's unknown radius, was moved up for clearer comparison with
the data for 2MASS-1507 (in  black), taken from Kirkpatrick \etal
(1999a).  The model shown is for \teff = 1700 K and a gravity of
$10^{4.5}$ cm s$^{-2}$. The abundance for FeH was calculated using
the new thermochemical data derived in this paper and for CrH
using the thermochemical data derived in Burrows et al. (2002) and
were not adjusted. The new solar abundances from Allende-Prieto et
al. (2002) for oxygen and carbon were used and this has reduced to
an acceptable level the depth of the water feature around 0.94
\mic. The effects of a forsterite cloud with particles of 50 \mic
radius incorporating 10\% of the available magnesium was included,
but had only a modest influence in this spectral range. Indicated
with arrows are the positions of the CrH, FeH, and H$_2$O
features. Also prominent are the Cs I lines at 8523 \AA\ and 8946
\AA. The resolution of the theoretical spectrum has been reduced
so that $R$($\lambda/{\Delta\lambda}$) is 1000. \label{fig:3}}
\end{figure}

Figure \ref{fig:3} portrays a comparison between the L5 dwarf 2MASS-1507 (Kirkpatrick \etal 1999a)
and a solar-metallicity theoretical spectrum at 0.85 \mic to $\sim$1.0 \mic,
calculated using the new FeH and CrH (Burrows \etal 2002)
opacities and abundances.  An effective temperature of 1700 K and a gravity of
10$^{4.5}$ cm s$^{-2}$ were assumed and the new solar elemental abundances from Allende-Prieto
et al. (2002) for oxygen and carbon were used.  The latter decrease the depth of the
water feature near 0.94 \mic and lead to a better fit in that
region of the spectrum than could previously be obtained.  The theoretical model was offset
for clarity of comparison with the data.  Since 2MASS-1507 is an L dwarf,
silicate clouds figure prominently in their atmospheres (Burrows \etal 2001).
We have incorporated a forsterite cloud with 50-\mic particles that sequesters 10\% of the solar magnesium.
Such a cloud, while having an important effect in the $J$, $H$, and $K$ bands, does
not dominate the depicted spectrum.  Clearly, the FeH and CrH features
around the Wing-Ford band and near the 0.86/0.87 \mic features are reasonably reproduced, along
with the overall spectral slope.  The fit is improved just longward of 0.87 \mic due to
the inclusion of the FeH opacities redward of the band head.  The relative strengths
of the paired FeH and CrH features depicted seem good, though playing with
the abundances, effective temperature, cloud model, and gravity could no doubt
further improve the fit.

Figure \ref{fig:3} demonstrates improvement over previous fits, but merely
represents the efforts that can and will be undertaken in the future as new, higher
resolution observations which highlight regions of the spectrum in which FeH plays
an important role are obtained.  In this spirit, we plan to incorporate the new FeH opacities derived in this
paper into a future, more comprehensive, paper that studies the defining FeH and CrH features observed in L dwarf
spectra, in particular in the $I$, $Z$, $J$, and $H$ photometric bands.

\section{ACKNOWLEDGMENTS}

This work was supported in part by NASA under grants
NAG5-10760 and NAG5-10629. Support
was also provided by the NASA Laboratory Physics Program and the
Natural Sciences and Engineering Research Council of Canada.
The authors would like to thank Richard Freedman for providing
guidance with line broadening parameters for FeH.

\clearpage

\tabletypesize{\scriptsize}
\begin{center}
\begin{tabular}{*{5}{c@{\hspace{0.5in}}}c}
\multicolumn{5}{c}{Table 1. Summary of IC-MRCI+Q spectroscopic constants, in cm$^{-1}$.}\\
\tableline\tableline
& $B_0$ & $\omega_e$ & $\omega_e x_e$ & $T_0$\\
\multispan 5 \hfil Ground state  \hfil \\
IC-MRCI+Q &  6.6880& 1792.8& 35.9& \\
Expt$^a$ &   6.50907& 1831.80& 34.86\\
\multispan 5 \hfil Excited state  \hfil \\
IC-MRCI+Q & 5.7019& 1543.3& 31.4& 9219.0\\
Expt   &  5.86663& 1501.55& 37.80& 9995.76\\
\noalign{\vskip 10pt}
\multispan 5 \hrulefill \\
\noalign{\vskip 10pt}

\multispan 5 \hfil Shifted potentials \hfil \\
& $B_0$ & $B_1$ & $B_2$\\
Ground State &  6.5092& 6.1359& 5.7809\\
Excited State&  5.8667& 5.5071& 5.1690\\
\tableline
\end{tabular}
\end{center}
$^a$ Present work, averaged over the lambda doublet components.

\clearpage

\begin{center}
Table 2. The Einstein A values, Franck-Condon factors, and
energy differences.
\begin{tabular}{rrrrrr}
\tableline\tableline
$v^\prime$ & $v^{\prime \prime}$ & FC & $\Delta$E(cm$^{-1}$)& A(s$^{-1}$) \\

0&     0&     0.8338&     9995.8&  0.1018(+7)\\
0&     1&     0.1616&     8226.2&  0.9371(+5)\\
0&     2&     0.0045&     6526.4&  0.8730(+3)\\
0&     3&     0.0001&     4898.7&  0.9800(+1)\\
0&     4&     0.0000&     3345.0&  0.1529(+0)\\
1&     0&     0.1349&    11422.3&  0.2772(+6)\\
1&     1&     0.5471&     9652.7&  0.5721(+6)\\
1&     2&     0.3015&     7952.9&  0.1506(+6)\\
1&     3&     0.0159&     6325.2&  0.2669(+4)\\
1&     4&     0.0005&     4771.5&  0.3059(+2)\\
2&     0&     0.0234&    12771.0&  0.7185(+5)\\
2&     1&     0.2104&    11001.5&  0.3724(+6)\\
2&     2&     0.3199&     9301.6&  0.2808(+6)\\
2&     3&     0.4082&     7673.9&  0.1728(+6)\\
2&     4&     0.0366&     6120.2&  0.5135(+4)\\
3&     0&     0.0053&    14041.6&  0.2224(+5)\\
3&     1&     0.0558&    12272.0&  0.1481(+6)\\
3&     2&     0.2365&    10572.2&  0.3569(+6)\\
3&     3&     0.1563&     8944.5&  0.1112(+6)\\
3&     4&     0.4741&     7390.8&  0.1665(+6)\\
4&     0&     0.0015&    15234.2&  0.8446(+4)\\
4&     1&     0.0159&    13464.7&  0.5823(+5)\\
4&     2&     0.0876&    11764.8&  0.1983(+6)\\
4&     3&     0.2256&    10137.1&  0.2873(+6)\\
4&     4&     0.0541&     8583.4&  0.2830(+5)\\
\tableline
\end{tabular}
\end{center}
\noindent

\clearpage
\begin{center}

\begin{tabular}{*{5}{c@{\hspace{0.5in}}}c}
\multicolumn{6}{c}{Table 3. ${^4\Delta} - {^4\Delta}$ H\"{o}nl-London Factors --
Hund's Case (a) Coupling}\\
\multicolumn{6}{c}{ }\\
Branch& $\Omega^{\prime\prime}$ & Parity & $\Omega^{\prime}$ & Parity & $S(J)$\\
\multicolumn{6}{c}{ }\\
$ P(J) $ & 7/2 & e/f & 7/2 & e/f & $\displaystyle{\frac{1}{4}\cdot
\frac{(2 J - 7) \: (2 J + 7)}{J}}$\\
\multicolumn{6}{c}{ }\\
$ P(J) $ & 5/2 & e/f & 5/2 & e/f & $\displaystyle{\frac{1}{4}\cdot
\frac{(2 J - 5) \: (2 J + 5)}{J}}$\\
\multicolumn{6}{c}{ }\\
$ P(J) $ & 3/2 & e/f & 3/2 & e/f & $\displaystyle{\frac{1}{4}\cdot
\frac{(2 J - 3) \: (2 J + 3)}{J}}$\\
\multicolumn{6}{c}{ }\\
$ P(J) $ & 1/2 & e/f & 1/2 & e/f & $\displaystyle{\frac{1}{4}\cdot
\frac{(2 J - 1) \: (2 J + 1)}{J}}$\\
\multicolumn{6}{c}{ }\\
$ Q(J) $ & 7/2 & e/f & 7/2 & f/e & $\displaystyle{\frac{49}{4}\cdot
\frac{2 J + 1}{J \: (J + 1)}}$\\
\multicolumn{6}{c}{ }\\
$ Q(J) $ & 5/2 & e/f & 5/2 & f/e & $\displaystyle{\frac{25}{4}\cdot
\frac{2 J + 1}{J \: (J + 1)}}$\\
\multicolumn{6}{c}{ }\\
$ Q(J) $ & 3/2 & e/f & 3/2 & f/e & $\displaystyle{\frac{9}{4}\cdot
\frac{2 J + 1}{J \: (J + 1)}}$\\
\multicolumn{6}{c}{ }\\
$ Q(J) $ & 1/2 & e/f & 1/2 & f/e & $\displaystyle{\frac{1}{4}\cdot
\frac{2 J + 1}{J \: (J + 1)}}$\\
\multicolumn{6}{c}{ }\\
$ R(J) $ & 7/2 & e/f & 7/2 & e/f & $\displaystyle{\frac{1}{4}\cdot
\frac{(2 J - 5) \: (2 J + 9)}{J + 1}}$\\
\multicolumn{6}{c}{ }\\
$ R(J) $ & 5/2 & e/f & 5/2 & e/f & $\displaystyle{\frac{1}{4}\cdot
\frac{(2 J - 3) \: (2 J + 7)}{J + 1}}$\\
\multicolumn{6}{c}{ }\\
$ R(J) $ & 3/2 & e/f & 3/2 & e/f & $\displaystyle{\frac{1}{4}\cdot
\frac{(2 J - 1) \: (2 J + 5)}{J + 1}}$\\
\multicolumn{6}{c}{ }\\
$ R(J) $ & 1/2 & e/f & 1/2 & e/f & $\displaystyle{\frac{1}{4}\:
\frac{(2 J + 1) \: (2 J + 3)}{J + 1}}$\\
\multicolumn{6}{c}{ }\\
\end{tabular}
\end{center}

\newpage

\begin{center}
\begin{tabular}{*{5}{c@{\hspace{12pt}}}c}
\multicolumn{6}{c}{Table 4. ${^4\Delta} - {^4\Delta}$ H\"{o}nl-London Factors --
Hund's Case (b) Coupling}\\
\multicolumn{6}{c}{ }\\
Branch& $ N^{\prime\prime}$ & Parity & $N^{\prime}$ & Parity & $S(J)$\\
\multicolumn{6}{c}{ }\\
$ ^PP(J) $ & $J - 3/2$ & $+/-$ & $J - 5/2$ & $+/-$ &
$\displaystyle{\frac{1}{4}\cdot\frac{(2 J - 7)\:(2 J + 1)^2}
{(2 J - 3)\:(J - 1)}}$\\
\multicolumn{6}{c}{ }\\
$ ^QP(J) $ & $J - 3/2$ & $+/-$ & $J - 3/2$ & $+/-$ &
$\displaystyle{48\cdot\frac{2 J + 1}
{J\:( 2 J - 3)\:(2 J - 1)^2}}$\\
\multicolumn{6}{c}{ }\\
$ ^RP(J) $ & $J - 3/2$ & $+/-$ & $J - 1/2$ & $+/-$ &
$\displaystyle{\frac{3}{4}\cdot \frac{(2 J - 5)\:(2 J + 3)}
{J^2\:(2 J - 1)^2\:(J - 1)}}$\\
\multicolumn{6}{c}{ }\\
$ ^PP(J) $ & $J - 1/2$ & $+/-$ & $J - 3/2$ & $+/-$ &
$\displaystyle{\frac{1}{4}\cdot \frac{(2 J - 5)\:(2 J - 3)\:(2 J +
1)\:(2 J + 3)\:(J + 1)}
{J^2\:(2 J - 1)^2}}$\\
\multicolumn{6}{c}{ }\\
$ ^QP(J) $ & $J - 1/2$ & $+/-$ & $J - 1/2$ & $+/-$ &
$\displaystyle{256\cdot \frac{(J - 1)\:(J + 1)}
{J\:(2 J - 1)^2\:(2 J + 1)^2}}$\\
\multicolumn{6}{c}{ }\\
$ ^RP(J) $ & $J - 1/2$ & $+/-$ & $J + 1/2$ & $+/-$ &
$\displaystyle{\frac{3}{4}\cdot \frac{(2 J - 3)\:(2 J + 5)}
{J^2\:(2 J + 1)^2\:(J + 1)}}$\\
\multicolumn{6}{c}{ }\\
$ ^PP(J) $ & $J + 1/2$ & $+/-$ & $J - 1/2$ & $+/-$ &
$\displaystyle{\frac{1}{4}\cdot \frac{(2 J - 3)\:(2 J - 1)\:(2 J +
3)\:(2 J + 5)\:(J - 1)}
{J^2\:(2 J + 1)^2}}$\\
\multicolumn{6}{c}{ }\\
$ ^QP(J) $ & $J + 1/2$ & $+/-$ & $J + 1/2$ & $+/-$ &
$\displaystyle{48\cdot \frac{2 J - 1}
{J\:(2 J + 1)^2\:(2 J + 3)}}$\\
\multicolumn{6}{c}{ }\\
$ ^PP(J) $ & $J + 3/2$ & $+/-$ & $J + 1/2$ & $+/-$ &
$\displaystyle{\frac{1}{4}\cdot \frac{(2 J - 1)^2\:(2 J + 7)}
{(2 J + 3)\:(J + 1)}}$\\
\multicolumn{6}{c}{ }\\
$ ^QQ(J) $ & $J - 3/2$ & $+/-$ & $J - 3/2$ & $+/-$ &
$\displaystyle{16\cdot \frac{(2 J + 1)\:(J + 1)}
{J\:(2 J - 1)^2}}$\\
\multicolumn{6}{c}{ }\\
$ ^RQ(J) $ & $J - 3/2$ & $+/-$ & $J - 1/2$ & $+/-$ &
$\displaystyle{\frac{3}{4}\cdot \frac{(2 J - 5)\:(2 J + 1)\:(2 J +
3)}
{J^2\:(2 J - 1)^2}}$\\
\multicolumn{6}{c}{ }\\
$ ^PQ(J) $ & $J - 1/2$ & $+/-$ & $J - 3/2$ & $+/-$ &
$\displaystyle{\frac{3}{4}\cdot \frac{(2 J - 5)\:(2 J + 1)\:(2 J +
3)}
{J^2\:(2 J - 1)^2}}$\\
\multicolumn{6}{c}{ }\\
$ ^QQ(J) $ & $J - 1/2$ & $+/-$ & $J - 1/2$ & $+/-$ &
$\displaystyle{16\cdot \frac{(2 J^2 + J - 4)^2}
{J\:(2 J - 1)^2\:(2 J + 1)\:(J + 1)}}$\\
\multicolumn{6}{c}{ }\\
$ ^RQ(J) $ & $J - 1/2$ & $+/-$ & $J + 1/2$ & $+/-$ &
$\displaystyle{\frac{1}{4}\cdot \frac{(2 J - 3)\:(2 J - 1)\:(2 J +
3)\:(2 J + 5)}
{J^2\:(2 J + 1)\:(J + 1)^2}}$\\
\multicolumn{6}{c}{ }\\
\end{tabular}
\end{center}

\newpage

\begin{center}
\begin{tabular}{*{5}{c@{\hspace{12pt}}}c}
\multicolumn{6}{c}{Table 4. ${^4\Delta} - {^4\Delta}$
H\"{o}nl-London Factors --
Hund's Case (b) Coupling}\\
\multicolumn{6}{c}{Continued}\\
\multicolumn{6}{c}{ }\\
$ ^PQ(J) $ & $J + 1/2$ & $+/-$ & $J - 1/2$ & $+/-$ &
$\displaystyle{\frac{1}{4}\cdot \frac{(2 J - 3)\:(2 J - 1)\:(2 J +
3)\:(2 J + 5)}
{J^2\:(2 J + 1)\:(J + 1)^2}}$\\
\multicolumn{6}{c}{ }\\
$ ^QQ(J) $ & $J + 1/2$ & $+/-$ & $J + 1/2$ & $+/-$ &
$\displaystyle{16\cdot \frac{(2 J^2 + 3 J - 3)^2}
{J\:(2 J + 1)\:(2 J + 3)^2\:(J + 1)}}$\\
\multicolumn{6}{c}{ }\\
$ ^RQ(J) $ & $J + 1/2$ & $+/-$ & $J + 3/2$ & $+/-$ &
$\displaystyle{\frac{3}{4}\cdot \frac{(2 J - 1)\:(2 J + 1)\:(2 J +
7)}
{(2 J + 3)^2\:(J + 1)^2}}$\\
\multicolumn{6}{c}{ }\\
$ ^PQ(J) $ & $J + 3/2$ & $+/-$ & $J + 1/2$ & $+/-$ &
$\displaystyle{\frac{3}{4}\cdot \frac{(2 J - 1)\:(2 J + 1)\:(2 J +
7)}
{(2 J + 3)^2\:(J + 1)^2}}$\\
\multicolumn{6}{c}{ }\\
$ ^QQ(J) $ & $J + 3/2$ & $+/-$ & $J + 3/2$ & $+/-$ &
$\displaystyle{16\cdot \frac{J\:(2 J + 1)}
{(2 J + 3)^2\:(J + 1)}}$\\
\multicolumn{6}{c}{ }\\
$ ^RR(J) $ & $J - 3/2$ & $+/-$ & $J - 1/2$ & $+/-$ &
$\displaystyle{\frac{1}{4}\cdot \frac{(2 J - 5)\:(2 J + 3)^2}
{J\:(2 J - 1)}}$\\
\multicolumn{6}{c}{ }\\
$ ^QR(J) $ & $J - 1/2$ & $+/-$ & $J - 1/2$ & $+/-$ &
$\displaystyle{48\cdot \frac{2 J + 3}
{(2 J - 1)\:(2 J + 1)^2\:(J + 1)}}$\\
\multicolumn{6}{c}{ }\\
$ ^RR(J) $ & $J - 1/2$ & $+/-$ & $J + 1/2$ & $+/-$ &
$\displaystyle{\frac{1}{4}\cdot \frac{(2 J - 3)\:(2 J - 1)\:(2 J +
3)\:(2 J + 5)\:(J + 2)}
{(2 J + 1)^2\:(J + 1)^2}}$\\
\multicolumn{6}{c}{ }\\
$ ^PR(J) $ & $J + 1/2$ & $+/-$ & $J - 1/2$ & $+/-$ &
$\displaystyle{\frac{3}{4}\cdot \frac{(2 J - 3)\:(2 J + 5)}
{J\:(2 J + 1)^2\:(J + 1)^2}}$\\
\multicolumn{6}{c}{ }\\
$ ^QR(J) $ & $J + 1/2$ & $+/-$ & $J + 1/2$ & $+/-$ &
$\displaystyle{256\cdot \frac{J\:(J + 2)}
{(2 J + 1)^2\:(2 J + 3)^2\:(J + 1)}}$\\
\multicolumn{6}{c}{ }\\
$ ^RR(J) $ & $J + 1/2$ & $+/-$ & $J + 3/2$ & $+/-$ &
$\displaystyle{\frac{1}{4}\cdot \frac{J\:(2 J - 1)\:(2 J + 1)\:(2
J + 5)\:(2 J + 7)}
{(2 J + 3)^2\:(J + 1)^2}}$\\
\multicolumn{6}{c}{ }\\
$ ^PR(J) $ & $J + 3/2$ & $+/-$ & $J + 1/2$ & $+/-$ &
$\displaystyle{\frac{3}{4}\cdot \frac{(2 J - 1)\:(2 J + 7)}
{(2 J + 3)^2\:(J + 1)^2\:(J + 2)}}$\\
\multicolumn{6}{c}{ }\\
$ ^QR(J) $ & $J + 3/2$ & $+/-$ & $J + 3/2$ & $+/-$ &
$\displaystyle{48\cdot \frac{2 J + 1}
{(J + 1)\:(2 J + 3)^2\:(2 J + 5)}}$\\
\multicolumn{6}{c}{ }\\
$ ^RR(J) $ & $J + 3/2$ & $+/-$ & $J + 5/2$ & $+/-$ &
$\displaystyle{\frac{1}{4}\cdot \frac{(2 J + 1)^2\:(2 J + 9)}
{(2 J + 5)\:(J + 2)}}$\\
\multicolumn{6}{c}{ }\\
\end{tabular}
\end{center}

\clearpage

\begin{center}
{Table 5. $^4\Delta$ Hund's Case (a) Hamiltonian Matrix}\\
{\ }\\
{$x = J + 1/2$}\\
{\ }\\
\end{center}
\begin{eqnarray*}
<7/2|H|7/2> &=& T_0+3A+2\lambda+B(x^2-7)-D(x^4-11x^2+22)\\[5pt]
<7/2|H|5/2> &=& <5/2|H|7/2> \:=\: -(B-2D(x^2-3))\sqrt{3(x^2-9)}\\[5pt]
<7/2|H|3/2> &=& <3/2|H|7/2> \:=\: -2D\sqrt{3(x^2-9)(x^2-4)}\\[5pt]
<7/2|H|1/2> &=& <1/2|H|7/2> \:=\: 0\\[15pt]
<5/2|H|5/2> &=& T_0+A-2\lambda+B(x^2+1)-D(x^4+9x^2-42)\\[5pt]
<5/2|H|3/2> &=& <3/2|H|5/2> \:=\: -2(B-2D(x^2+3))\sqrt{x^2-4}\\[5pt]
<5/2|H|1/2> &=& <1/2|H|5/2> \:=\: -2D\sqrt{3(x^2-4)(x^2-1)}\\[15pt]
<3/2|H|3/2> &=& T_0-A-2\lambda+B(x^2+5)-D(x^4+17x^2+6)\\[5pt]
<3/2|H|1/2> &=& <1/2|H|3/2> \:=\: -(B-2D(x^2+5))\sqrt{3(x^2-1)}\\[15pt]
<1/2|H|1/2> &=& T_0-3A+2\lambda+B(x^2+5)-D(x^4+13x^2+22)
\end{eqnarray*}

\clearpage

\begin{center}
{Table 6. $^4\Delta$ Hund's Case (b) Hamiltonian Matrix}\\
{\ }\\
\begin{tabular}{l}
\multicolumn{1}{c}{$x = N = J - 3/2$}\\[10pt]
$\displaystyle{<x|H|x> = T + B x(x+1) - D x^2(x+1)^2 + \frac{6 A}{x+1}
-2\lambda\frac{(x-3)(x+4)}{(x+1)(2x+3)}}$\\[15pt]
$\displaystyle{<x|H|x+1> = <x+1|H|x> \:=\: (A(x+2)+4\lambda)\sqrt{
\frac{3(x-1)(x+3)(2x+5)}{(x+1)^2(x+2)^2(2x+3)}}}$\\[15pt]
$\displaystyle{<x|H|x+2> = <x+2|H|x> \:=\: 2\lambda\sqrt{
\frac{3(x-1)x(x+3)^2(x+4)}{(x+1)(x+2)^2(2x+3)^2}}}$\\[15pt]
$\displaystyle{<x|H|x+3> = <x+3|H|x> \:=\: 0}$\\[25pt]
\multicolumn{1}{c}{$x = N = J - 1/2$}\\[10pt]
$\displaystyle{<x|H|x> = T + B x(x+1) - D x^2(x+1)^2 + 2A\frac{x-3}{x(x+1)}
+2\lambda\frac{(x-3)(x+3)(x+4)}{x(x+1)(2x+3)}}$\\[15pt]
$\displaystyle{<x|H|x+1> = <x+1|H|x> \:=\: 4(Ax(x+2)-6\lambda)\sqrt{
\frac{(x-1)(x+3)}{x(x+1)^2(x+2)(2x+1)(2x+3)}}}$\\[15pt]
$\displaystyle{<x|H|x+2> = <x+2|H|x> \:=\: 2\lambda\sqrt{
\frac{3(x-1)x^2(x+3)(x+4)}{(x+1)^2(x+2)(2x+3)^2}}}$\\[25pt]
\multicolumn{1}{c}{$x = N = J + 1/2$}\\[10pt]
$\displaystyle{<x|H|x> = T + B x(x+1) - D x^2(x+1)^2 - 2A\frac{x+4}{x(x+1)}
+2\lambda\frac{(x-3)(x-2)(x+4)}{x(x+1)(2x-1)}}$\\[15pt]
$\displaystyle{<x|H|x+1> = <x+1|H|x> \:=\: (Ax-4\lambda)\sqrt{
\frac{3(x-1)(x+3)(2x-1)}{x^2(x+1)^2(2x+1)}}}$\\[25pt]
\multicolumn{1}{c}{$x = N = J + 3/2$}\\[10pt]
$\displaystyle{<x|H|x> = T + B x(x+1) - D x^2(x+1)^2 - \frac{6A}{x}
-2\lambda\frac{(x-3)(x+4)}{x(2x-1)}}$\\
\end{tabular}
\end{center}

\newpage

\begin{center}
\begin{tabular}{l@{\hspace{12pt}}*{3}{r@{\hspace{0in}}c@{\hspace{0in}}l}}
\multicolumn{10}{c}{Table 7. Molecular Constants for the $X^4\Delta$ and $F^4\Delta$ States}\\
{\ }\\
\multicolumn{10}{c}{$X^4\Delta$ State}\\
$A$            &   --116&.&860(524) & \multicolumn{6}{c}{ }\\
$\lambda$      &      10&.&645(693) & \multicolumn{6}{c}{ }\\
\multicolumn{10}{c}{Parity a}\\
  & \multicolumn{3}{c}{$v = 0$} & \multicolumn{3}{c}{$v = 1$}
  & \multicolumn{3}{c}{$v = 2$}\\
$T_v$ & 315&.&879(221) & 2078&.&171(358) & 3769&.&400(361)\\
$B_v$ & 6&.&49504(312) & 6&.&29867(516)  & 6&.&16724(471)\\
$D_v$ & 0&.&00022624(833) & 0&.&0000592(139) & 0&.&0001586(136)\\
$\omega_e$     &  1833&.&36(113)    & \multicolumn{6}{c}{ }\\
$\omega_ex_e$  &    35&.&53(50)     & \multicolumn{6}{c}{ }\\
$B_e$          &     6&.&5662(185)  & \multicolumn{6}{c}{ }\\
$\alpha_e$     &     0&.&1639(108)  & \multicolumn{6}{c}{ }\\
\multicolumn{10}{c}{Parity b}\\
  & \multicolumn{3}{c}{$v = 0$} & \multicolumn{3}{c}{$v = 1$}
  & \multicolumn{3}{c}{$v = 2$}\\
$T_v$ & 319&.&5395(458) & 2081&.&421(627) & 3774&.&95(134)\\
$B_v$ & 6&.&523101(590) & 6&.&34794(918)  & 6&.&1016(226)\\
$D_v$ & 0&.&00023866(143) & 0&.&0001967(261) & --0&.&0001767(800)\\
$\omega_e$     &  1830&.&23(189)    & \multicolumn{6}{c}{ }\\
$\omega_ex_e$  &    34&.&18(80)     & \multicolumn{6}{c}{ }\\
$B_e$          &     6&.&6403(203)  & \multicolumn{6}{c}{ }\\
$\alpha_e$     &     0&.&2108(119)  & \multicolumn{6}{c}{ }\\
\end{tabular}
\end{center}

\begin{center}
\begin{tabular}{l@{\hspace{12pt}}*{3}{r@{\hspace{0in}}c@{\hspace{0in}}l}}
\multicolumn{10}{c}{Table 7. Molecular Constants for the $X^4\Delta$ and $F^4\Delta$ States, Continued}\\
{\ }\\
\multicolumn{10}{c}{$F^4\Delta$ State}\\
$A$            &   --124&.&816(692) & \multicolumn{6}{c}{ }\\
$\lambda$      &     --9&.&263(916) & \multicolumn{6}{c}{ }\\
\multicolumn{10}{c}{Parity a}\\
  & \multicolumn{3}{c}{$v = 0$} & \multicolumn{3}{c}{$v = 1$}
  & \multicolumn{3}{c}{$v = 2$}\\
$T_v$ & 10313&.&114(186) & 11737&.&011(440) & 13085&.&964(591)\\
$B_v$ &  5&.&87584(210)  &  5&.&69283(553)  &  5&.&45930(847)\\
$D_v$ &  0&.&00040430(421) & 0&.&0004711(138) & 0&.&0004365(249)\\
$\omega_e$     &  1498&.&84(145)    & \multicolumn{6}{c}{ }\\
$\omega_ex_e$  &    37&.&47(60)     & \multicolumn{6}{c}{ }\\
$B_e$          &     5&.&9884(147)  & \multicolumn{6}{c}{ }\\
$\alpha_e$     &     0&.&20827(842) & \multicolumn{6}{c}{ }\\
\multicolumn{10}{c}{Parity b}\\
  & \multicolumn{3}{c}{$v = 0$} & \multicolumn{3}{c}{$v = 1$}
  & \multicolumn{3}{c}{$v = 2$}\\
$T_v$ & 10313&.&819(225) & 11741&.&817(195) & 13093&.&571(481)\\
$B_v$ &   5&.&85742(290) &   5&.&60903(264) & 5&.&38113(718)\\
$D_v$ & 0&.&00032849(704) & 0&.&00034484(684) & 0&.&0002624(204)\\
$\omega_e$     &  1504&.&24(80)     & \multicolumn{6}{c}{ }\\
$\omega_ex_e$  &    38&.&12(40)     & \multicolumn{6}{c}{ }\\
$B_e$          &     5&.&97308(583) & \multicolumn{6}{c}{ }\\
$\alpha_e$     &     0&.&23815(342) & \multicolumn{6}{c}{ }\\
\end{tabular}
\end{center}

\newpage

\begin{center}
\begin{tabular}{*{9}{@{\hspace{12pt}}r}}
\multicolumn{9}{c}{Table 8a. The $X^4\Delta$ $v$ = 0 Term Values}\\
{\ }\\
$J$ & \multicolumn{2}{c}{7/2} & \multicolumn{2}{c}{5/2} &
 \multicolumn{2}{c}{3/2} & \multicolumn{2}{c}{1/2}\\
    &\multicolumn{1}{c}{a} &\multicolumn{1}{c}{b}
    &\multicolumn{1}{c}{a} &\multicolumn{1}{c}{b}
    &\multicolumn{1}{c}{a} &\multicolumn{1}{c}{b}
    &\multicolumn{1}{c}{a} &\multicolumn{1}{c}{b}\\
    &\multicolumn{2}{c}{$\overbrace{\mbox{\hspace{80pt}}}$}
    &\multicolumn{2}{c}{$\overbrace{\mbox{\hspace{80pt}}}$}
    &\multicolumn{2}{c}{$\overbrace{\mbox{\hspace{80pt}}}$}
    &\multicolumn{2}{c}{$\overbrace{\mbox{\hspace{80pt}}}$}\\
 0.5&          &          &          &          &          &          &    694.40&    699.70\\
 1.5&          &          &          &          &    450.60&    450.92&    716.44&    723.28\\
 2.5&          &          &    244.68&    244.68&    484.89&    486.22&    753.10&    762.42\\
\end{tabular}
\end{center}

\begin{center}
\begin{tabular}{*{9}{@{\hspace{12pt}}r}}
\multicolumn{9}{c}{Table 8a. The $X^4\Delta$ $v$ = 0 Term Values, Continued}\\
{\ }\\
$J$ & \multicolumn{2}{c}{7/2} & \multicolumn{2}{c}{5/2} &
 \multicolumn{2}{c}{3/2} & \multicolumn{2}{c}{1/2}\\
    &\multicolumn{1}{c}{a} &\multicolumn{1}{c}{b}
    &\multicolumn{1}{c}{a} &\multicolumn{1}{c}{b}
    &\multicolumn{1}{c}{a} &\multicolumn{1}{c}{b}
    &\multicolumn{1}{c}{a} &\multicolumn{1}{c}{b}\\
    &\multicolumn{2}{c}{$\overbrace{\mbox{\hspace{80pt}}}$}
    &\multicolumn{2}{c}{$\overbrace{\mbox{\hspace{80pt}}}$}
    &\multicolumn{2}{c}{$\overbrace{\mbox{\hspace{80pt}}}$}
    &\multicolumn{2}{c}{$\overbrace{\mbox{\hspace{80pt}}}$}\\
36.5&   8013.02&   7940.01&   8355.88&   8390.47&   8928.34&   8917.64&   9525.92&   9568.05\\
37.5&   8438.71&   8360.51&   8781.51&   8816.47&   9365.55&   9352.86&   9975.33&  10017.87\\
38.5&   8873.65&   8790.00&   9216.06&   9251.26&   9811.63&   9796.75&  10433.58&  10476.37\\
\end{tabular}
\end{center}

\newpage

\begin{center}
\begin{tabular}{*{9}{@{\hspace{12pt}}r}}
\multicolumn{9}{c}{Table 8b. The $X^4\Delta$ $v$ = 1 Term Values}\\
{\ }\\
$J$ & \multicolumn{2}{c}{7/2} & \multicolumn{2}{c}{5/2} &
 \multicolumn{2}{c}{3/2} & \multicolumn{2}{c}{1/2}\\
    &\multicolumn{1}{c}{a} &\multicolumn{1}{c}{b}
    &\multicolumn{1}{c}{a} &\multicolumn{1}{c}{b}
    &\multicolumn{1}{c}{a} &\multicolumn{1}{c}{b}
    &\multicolumn{1}{c}{a} &\multicolumn{1}{c}{b}\\
    &\multicolumn{2}{c}{$\overbrace{\mbox{\hspace{80pt}}}$}
    &\multicolumn{2}{c}{$\overbrace{\mbox{\hspace{80pt}}}$}
    &\multicolumn{2}{c}{$\overbrace{\mbox{\hspace{80pt}}}$}
    &\multicolumn{2}{c}{$\overbrace{\mbox{\hspace{80pt}}}$}\\
 0.5&          &          &          &          &          &          &   2454.80&   2463.00\\
 1.5&          &          &          &          &   2203.40&   2210.20&   2475.70&   2485.20\\
 2.5&          &          &   2005.10&   2005.60&   2238.80&   2245.00&   2511.82&   2522.10\\
\end{tabular}
\end{center}

\begin{center}
\begin{tabular}{*{9}{@{\hspace{12pt}}r}}
\multicolumn{9}{c}{Table 8b. The $X^4\Delta$ $v$ = 1 Term Values, Continued}\\
{\ }\\
$J$ & \multicolumn{2}{c}{7/2} & \multicolumn{2}{c}{5/2} &
 \multicolumn{2}{c}{3/2} & \multicolumn{2}{c}{1/2}\\
    &\multicolumn{1}{c}{a} &\multicolumn{1}{c}{b}
    &\multicolumn{1}{c}{a} &\multicolumn{1}{c}{b}
    &\multicolumn{1}{c}{a} &\multicolumn{1}{c}{b}
    &\multicolumn{1}{c}{a} &\multicolumn{1}{c}{b}\\
    &\multicolumn{2}{c}{$\overbrace{\mbox{\hspace{80pt}}}$}
    &\multicolumn{2}{c}{$\overbrace{\mbox{\hspace{80pt}}}$}
    &\multicolumn{2}{c}{$\overbrace{\mbox{\hspace{80pt}}}$}
    &\multicolumn{2}{c}{$\overbrace{\mbox{\hspace{80pt}}}$}\\
36.5&   9719.98&   9553.94&  10257.88&  10100.94&  10820.90&  10636.94&  11400.49&  11208.42\\
37.5&  10159.76&   9971.16&  10708.84&  10528.59&  11283.86&  11074.19&  11875.77&  11656.29\\
38.5&  10610.97&  10397.72&  11171.26&  10965.52&  11758.32&  11520.64&  12362.65&  12113.30\\
\end{tabular}
\end{center}

\newpage

\begin{center}
\begin{tabular}{*{9}{@{\hspace{12pt}}r}}
\multicolumn{9}{c}{Table 8c. The $X^4\Delta$ $v$ = 2 Term Values}\\
{\ }\\
$J$ & \multicolumn{2}{c}{7/2} & \multicolumn{2}{c}{5/2} &
 \multicolumn{2}{c}{3/2} & \multicolumn{2}{c}{1/2}\\
    &\multicolumn{1}{c}{a} &\multicolumn{1}{c}{b}
    &\multicolumn{1}{c}{a} &\multicolumn{1}{c}{b}
    &\multicolumn{1}{c}{a} &\multicolumn{1}{c}{b}
    &\multicolumn{1}{c}{a} &\multicolumn{1}{c}{b}\\
    &\multicolumn{2}{c}{$\overbrace{\mbox{\hspace{80pt}}}$}
    &\multicolumn{2}{c}{$\overbrace{\mbox{\hspace{80pt}}}$}
    &\multicolumn{2}{c}{$\overbrace{\mbox{\hspace{80pt}}}$}
    &\multicolumn{2}{c}{$\overbrace{\mbox{\hspace{80pt}}}$}\\
 0.5&          &          &          &          &          &          &   4187.20&   4174.50\\
 1.5&          &          &          &          &   3909.10&   3903.80&   4205.40&   4193.60\\
 2.5&          &          &   3698.50&   3693.90&   3940.10&   3937.20&   4236.00&   4225.50\\
\end{tabular}
\end{center}

\begin{center}
\begin{tabular}{*{9}{@{\hspace{12pt}}r}}
\multicolumn{9}{c}{Table 8c. The $X^4\Delta$ $v$ = 2 Term Values, Continued}\\
{\ }\\
$J$ & \multicolumn{2}{c}{7/2} & \multicolumn{2}{c}{5/2} &
 \multicolumn{2}{c}{3/2} & \multicolumn{2}{c}{1/2}\\
    &\multicolumn{1}{c}{a} &\multicolumn{1}{c}{b}
    &\multicolumn{1}{c}{a} &\multicolumn{1}{c}{b}
    &\multicolumn{1}{c}{a} &\multicolumn{1}{c}{b}
    &\multicolumn{1}{c}{a} &\multicolumn{1}{c}{b}\\
    &\multicolumn{2}{c}{$\overbrace{\mbox{\hspace{80pt}}}$}
    &\multicolumn{2}{c}{$\overbrace{\mbox{\hspace{80pt}}}$}
    &\multicolumn{2}{c}{$\overbrace{\mbox{\hspace{80pt}}}$}
    &\multicolumn{2}{c}{$\overbrace{\mbox{\hspace{80pt}}}$}\\
36.5&  11120.73&  11596.24&  11622.53&  12160.56&  12148.78&  12733.40&  12706.22&  13288.99\\
37.5&  11534.61&  12068.60&  12045.26&  12647.12&  12580.92&  13234.32&  13148.49&  13803.42\\
38.5&  11958.14&  12555.83&  12477.56&  13148.76&  13022.59&  13750.59&  13600.25&  14333.46\\
\end{tabular}
\end{center}

\newpage

\begin{center}
\begin{tabular}{*{9}{@{\hspace{12pt}}r}}
\multicolumn{9}{c}{Table 8d. The $X^4\Delta$ $v$ = 3 Term Values}\\
{\ }\\
$J$ & \multicolumn{2}{c}{7/2} & \multicolumn{2}{c}{5/2} &
 \multicolumn{2}{c}{3/2} & \multicolumn{2}{c}{1/2}\\
    &\multicolumn{1}{c}{a} &\multicolumn{1}{c}{b}
    &\multicolumn{1}{c}{a} &\multicolumn{1}{c}{b}
    &\multicolumn{1}{c}{a} &\multicolumn{1}{c}{b}
    &\multicolumn{1}{c}{a} &\multicolumn{1}{c}{b}\\
    &\multicolumn{2}{c}{$\overbrace{\mbox{\hspace{80pt}}}$}
    &\multicolumn{2}{c}{$\overbrace{\mbox{\hspace{80pt}}}$}
    &\multicolumn{2}{c}{$\overbrace{\mbox{\hspace{80pt}}}$}
    &\multicolumn{2}{c}{$\overbrace{\mbox{\hspace{80pt}}}$}\\
 0.5&          &          &          &          &          &          &   5810.83&   5820.76\\
 1.5&          &          &          &          &   5551.33&   5561.02&   5829.95&   5839.58\\
 2.5&          &          &   5321.99&   5331.64&   5582.09&   5591.32&   5861.81&   5870.93\\
\end{tabular}
\end{center}

\begin{center}
\begin{tabular}{*{9}{@{\hspace{12pt}}r}}
\multicolumn{9}{c}{Table 8d. The $X^4\Delta$ $v$ = 3 Term Values, Continued}\\
{\ }\\
$J$ & \multicolumn{2}{c}{7/2} & \multicolumn{2}{c}{5/2} &
 \multicolumn{2}{c}{3/2} & \multicolumn{2}{c}{1/2}\\
    &\multicolumn{1}{c}{a} &\multicolumn{1}{c}{b}
    &\multicolumn{1}{c}{a} &\multicolumn{1}{c}{b}
    &\multicolumn{1}{c}{a} &\multicolumn{1}{c}{b}
    &\multicolumn{1}{c}{a} &\multicolumn{1}{c}{b}\\
    &\multicolumn{2}{c}{$\overbrace{\mbox{\hspace{80pt}}}$}
    &\multicolumn{2}{c}{$\overbrace{\mbox{\hspace{80pt}}}$}
    &\multicolumn{2}{c}{$\overbrace{\mbox{\hspace{80pt}}}$}
    &\multicolumn{2}{c}{$\overbrace{\mbox{\hspace{80pt}}}$}\\
36.5&  12429.21&  12342.06&  12898.29&  12807.00&  13364.59&  13269.38&  13829.49&  13730.59\\
37.5&  12814.71&  12723.06&  13290.60&  13194.79&  13763.27&  13663.52&  14234.11&  14130.64\\
38.5&  13208.01&  13111.83&  13690.52&  13590.19&  14169.41&  14065.12&  14646.03&  14538.01\\
\end{tabular}
\end{center}

\newpage

\begin{center}
\begin{tabular}{*{9}{@{\hspace{12pt}}r}}
\multicolumn{9}{c}{Table 8e. The $X^4\Delta$ $v$ = 4 Term Values}\\
{\ }\\
$J$ & \multicolumn{2}{c}{7/2} & \multicolumn{2}{c}{5/2} &
 \multicolumn{2}{c}{3/2} & \multicolumn{2}{c}{1/2}\\
    &\multicolumn{1}{c}{a} &\multicolumn{1}{c}{b}
    &\multicolumn{1}{c}{a} &\multicolumn{1}{c}{b}
    &\multicolumn{1}{c}{a} &\multicolumn{1}{c}{b}
    &\multicolumn{1}{c}{a} &\multicolumn{1}{c}{b}\\
    &\multicolumn{2}{c}{$\overbrace{\mbox{\hspace{80pt}}}$}
    &\multicolumn{2}{c}{$\overbrace{\mbox{\hspace{80pt}}}$}
    &\multicolumn{2}{c}{$\overbrace{\mbox{\hspace{80pt}}}$}
    &\multicolumn{2}{c}{$\overbrace{\mbox{\hspace{80pt}}}$}\\
 0.5&          &          &          &          &          &          &   7358.96&   7376.28\\
 1.5&          &          &          &          &   7099.04&   7116.00&   7377.54&   7394.40\\
 2.5&          &          &   6869.61&   6886.50&   7128.95&   7145.19&   7408.48&   7424.57\\
\end{tabular}
\end{center}

\begin{center}
\begin{tabular}{*{9}{@{\hspace{12pt}}r}}
\multicolumn{9}{c}{Table 8e. The $X^4\Delta$ $v$ = 4 Term Values, Continued}\\
{\ }\\
$J$ & \multicolumn{2}{c}{7/2} & \multicolumn{2}{c}{5/2} &
 \multicolumn{2}{c}{3/2} & \multicolumn{2}{c}{1/2}\\
    &\multicolumn{1}{c}{a} &\multicolumn{1}{c}{b}
    &\multicolumn{1}{c}{a} &\multicolumn{1}{c}{b}
    &\multicolumn{1}{c}{a} &\multicolumn{1}{c}{b}
    &\multicolumn{1}{c}{a} &\multicolumn{1}{c}{b}\\
    &\multicolumn{2}{c}{$\overbrace{\mbox{\hspace{80pt}}}$}
    &\multicolumn{2}{c}{$\overbrace{\mbox{\hspace{80pt}}}$}
    &\multicolumn{2}{c}{$\overbrace{\mbox{\hspace{80pt}}}$}
    &\multicolumn{2}{c}{$\overbrace{\mbox{\hspace{80pt}}}$}\\
36.5&  13802.82&  13671.66&  14264.62&  14127.12&  14724.02&  14580.50&  15182.44&  15033.27\\
37.5&  14180.38&  14042.28&  14648.97&  14504.49&  15114.73&  14964.18&  15579.07&  15422.83\\
38.5&  14565.73&  14420.66&  15040.96&  14889.49&  15512.92&  15355.36&  15983.05&  15819.77\\
\end{tabular}
\end{center}

\newpage

\begin{center}
\begin{tabular}{*{9}{@{\hspace{12pt}}r}}
\multicolumn{9}{c}{Table 8f. The $F^4\Delta$ $v$ = 0 Term Values}\\
{\ }\\
$J$ & \multicolumn{2}{c}{7/2} & \multicolumn{2}{c}{5/2} &
 \multicolumn{2}{c}{3/2} & \multicolumn{2}{c}{1/2}\\
    &\multicolumn{1}{c}{a} &\multicolumn{1}{c}{b}
    &\multicolumn{1}{c}{a} &\multicolumn{1}{c}{b}
    &\multicolumn{1}{c}{a} &\multicolumn{1}{c}{b}
    &\multicolumn{1}{c}{a} &\multicolumn{1}{c}{b}\\
    &\multicolumn{2}{c}{$\overbrace{\mbox{\hspace{80pt}}}$}
    &\multicolumn{2}{c}{$\overbrace{\mbox{\hspace{80pt}}}$}
    &\multicolumn{2}{c}{$\overbrace{\mbox{\hspace{80pt}}}$}
    &\multicolumn{2}{c}{$\overbrace{\mbox{\hspace{80pt}}}$}\\
 0.5&          &          &          &          &          &          &  10680.34&  10679.93\\
 1.5&          &          &          &          &  10486.31&  10486.52&  10699.13&  10698.83\\
 2.5&          &          &  10266.50&  10266.56&  10515.79&  10515.88&  10730.35&  10729.74\\
\end{tabular}
\end{center}

\begin{center}
\begin{tabular}{*{9}{@{\hspace{12pt}}r}}
\multicolumn{9}{c}{Table 8f. The $F^4\Delta$ $v$ = 0 Term Values, Continued}\\
{\ }\\
$J$ & \multicolumn{2}{c}{7/2} & \multicolumn{2}{c}{5/2} &
 \multicolumn{2}{c}{3/2} & \multicolumn{2}{c}{1/2}\\
    &\multicolumn{1}{c}{a} &\multicolumn{1}{c}{b}
    &\multicolumn{1}{c}{a} &\multicolumn{1}{c}{b}
    &\multicolumn{1}{c}{a} &\multicolumn{1}{c}{b}
    &\multicolumn{1}{c}{a} &\multicolumn{1}{c}{b}\\
    &\multicolumn{2}{c}{$\overbrace{\mbox{\hspace{80pt}}}$}
    &\multicolumn{2}{c}{$\overbrace{\mbox{\hspace{80pt}}}$}
    &\multicolumn{2}{c}{$\overbrace{\mbox{\hspace{80pt}}}$}
    &\multicolumn{2}{c}{$\overbrace{\mbox{\hspace{80pt}}}$}\\
36.5&  16978.84&  17082.96&  17362.27&  17468.14&  17756.77&  17871.08&  18163.72&  18298.94\\
37.5&  17327.90&  17445.13&  17713.35&  17833.08&  18110.80&  18239.97&  18521.24&  18672.65\\
38.5&  17682.07&  17813.60&  18069.30&  18204.14&  18469.42&  18614.75&  18883.02&  19051.96\\
\end{tabular}
\end{center}

\newpage

\begin{center}
\begin{tabular}{*{9}{@{\hspace{12pt}}r}}
\multicolumn{9}{c}{Table 8g. The $F^4\Delta$ $v$ = 1 Term Values}\\
{\ }\\
$J$ & \multicolumn{2}{c}{7/2} & \multicolumn{2}{c}{5/2} &
 \multicolumn{2}{c}{3/2} & \multicolumn{2}{c}{1/2}\\
    &\multicolumn{1}{c}{a} &\multicolumn{1}{c}{b}
    &\multicolumn{1}{c}{a} &\multicolumn{1}{c}{b}
    &\multicolumn{1}{c}{a} &\multicolumn{1}{c}{b}
    &\multicolumn{1}{c}{a} &\multicolumn{1}{c}{b}\\
    &\multicolumn{2}{c}{$\overbrace{\mbox{\hspace{80pt}}}$}
    &\multicolumn{2}{c}{$\overbrace{\mbox{\hspace{80pt}}}$}
    &\multicolumn{2}{c}{$\overbrace{\mbox{\hspace{80pt}}}$}
    &\multicolumn{2}{c}{$\overbrace{\mbox{\hspace{80pt}}}$}\\
 0.5&          &          &          &          &          &          &  12105.86&  12123.50\\
 1.5&          &          &          &          &  11878.20&  11913.60&  12124.08&  12137.50\\
 2.5&          &          &  11694.55&  11694.55&  11909.40&  11941.10&  12154.31&  12161.80\\
\end{tabular}
\end{center}

\begin{center}
\begin{tabular}{*{9}{@{\hspace{12pt}}r}}
\multicolumn{9}{c}{Table 8g. The $F^4\Delta$ $v$ = 1 Term Values, Continued}\\
{\ }\\
$J$ & \multicolumn{2}{c}{7/2} & \multicolumn{2}{c}{5/2} &
 \multicolumn{2}{c}{3/2} & \multicolumn{2}{c}{1/2}\\
    &\multicolumn{1}{c}{a} &\multicolumn{1}{c}{b}
    &\multicolumn{1}{c}{a} &\multicolumn{1}{c}{b}
    &\multicolumn{1}{c}{a} &\multicolumn{1}{c}{b}
    &\multicolumn{1}{c}{a} &\multicolumn{1}{c}{b}\\
    &\multicolumn{2}{c}{$\overbrace{\mbox{\hspace{80pt}}}$}
    &\multicolumn{2}{c}{$\overbrace{\mbox{\hspace{80pt}}}$}
    &\multicolumn{2}{c}{$\overbrace{\mbox{\hspace{80pt}}}$}
    &\multicolumn{2}{c}{$\overbrace{\mbox{\hspace{80pt}}}$}\\
36.5&  18052.78&  18174.76&  18422.04&  18548.06&  18793.89&  18932.28&  19192.07&  19329.22\\
37.5&  18375.87&  18516.15&  18746.22&  18891.78&  19119.83&  19279.31&  19520.47&  19680.18\\
38.5&  18702.64&  18863.09&  19073.79&  19240.84&  19448.83&  19631.45&  19851.53&  20035.95\\
\end{tabular}
\end{center}

\newpage

\begin{center}
\begin{tabular}{*{9}{@{\hspace{12pt}}r}}
\multicolumn{9}{c}{Table 8h. The $F^4\Delta$ $v$ = 2 Term Values}\\
{\ }\\
$J$ & \multicolumn{2}{c}{7/2} & \multicolumn{2}{c}{5/2} &
 \multicolumn{2}{c}{3/2} & \multicolumn{2}{c}{1/2}\\
    &\multicolumn{1}{c}{a} &\multicolumn{1}{c}{b}
    &\multicolumn{1}{c}{a} &\multicolumn{1}{c}{b}
    &\multicolumn{1}{c}{a} &\multicolumn{1}{c}{b}
    &\multicolumn{1}{c}{a} &\multicolumn{1}{c}{b}\\
    &\multicolumn{2}{c}{$\overbrace{\mbox{\hspace{80pt}}}$}
    &\multicolumn{2}{c}{$\overbrace{\mbox{\hspace{80pt}}}$}
    &\multicolumn{2}{c}{$\overbrace{\mbox{\hspace{80pt}}}$}
    &\multicolumn{2}{c}{$\overbrace{\mbox{\hspace{80pt}}}$}\\
 0.5&          &          &          &          &          &          &  13428.40&  13456.70\\
 1.5&          &          &          &          &  13231.50&  13273.60&  13448.30&  13473.90\\
 2.5&          &          &  13045.90&  13040.20&  13263.30&  13300.40&  13480.90&  13502.40\\
\end{tabular}
\end{center}

\begin{center}
\begin{tabular}{*{9}{@{\hspace{12pt}}r}}
\multicolumn{9}{c}{Table 8h. The $F^4\Delta$ $v$ = 2 Term Values, Continued}\\
{\ }\\
$J$ & \multicolumn{2}{c}{7/2} & \multicolumn{2}{c}{5/2} &
 \multicolumn{2}{c}{3/2} & \multicolumn{2}{c}{1/2}\\
    &\multicolumn{1}{c}{a} &\multicolumn{1}{c}{b}
    &\multicolumn{1}{c}{a} &\multicolumn{1}{c}{b}
    &\multicolumn{1}{c}{a} &\multicolumn{1}{c}{b}
    &\multicolumn{1}{c}{a} &\multicolumn{1}{c}{b}\\
    &\multicolumn{2}{c}{$\overbrace{\mbox{\hspace{80pt}}}$}
    &\multicolumn{2}{c}{$\overbrace{\mbox{\hspace{80pt}}}$}
    &\multicolumn{2}{c}{$\overbrace{\mbox{\hspace{80pt}}}$}
    &\multicolumn{2}{c}{$\overbrace{\mbox{\hspace{80pt}}}$}\\
36.5&  19151.19&  19346.23&  19522.62&  19736.66&  19898.95&  20136.24&  20275.78&  20541.86\\
37.5&  19463.94&  19686.48&  19836.72&  20080.26&  20215.15&  20484.25&  20594.37&  20894.85\\
38.5&  19780.45&  20033.14&  20154.34&  20430.15&  20534.56&  20838.37&  20915.80&  21253.75\\
\end{tabular}
\end{center}

\newpage

\begin{center}
\begin{tabular}{*{9}{@{\hspace{12pt}}r}}
\multicolumn{9}{c}{Table 8i. The $F^4\Delta$ $v$ = 3 Term Values}\\
{\ }\\
$J$ & \multicolumn{2}{c}{7/2} & \multicolumn{2}{c}{5/2} &
 \multicolumn{2}{c}{3/2} & \multicolumn{2}{c}{1/2}\\
    &\multicolumn{1}{c}{a} &\multicolumn{1}{c}{b}
    &\multicolumn{1}{c}{a} &\multicolumn{1}{c}{b}
    &\multicolumn{1}{c}{a} &\multicolumn{1}{c}{b}
    &\multicolumn{1}{c}{a} &\multicolumn{1}{c}{b}\\
    &\multicolumn{2}{c}{$\overbrace{\mbox{\hspace{80pt}}}$}
    &\multicolumn{2}{c}{$\overbrace{\mbox{\hspace{80pt}}}$}
    &\multicolumn{2}{c}{$\overbrace{\mbox{\hspace{80pt}}}$}
    &\multicolumn{2}{c}{$\overbrace{\mbox{\hspace{80pt}}}$}\\
 0.5&          &          &          &          &          &          &  14824.56&  14832.94\\
 1.5&          &          &          &          &  14626.58&  14634.66&  14841.49&  14849.46\\
 2.5&          &          &  14381.33&  14389.32&  14652.95&  14660.43&  14869.67&  14876.97\\
\end{tabular}
\end{center}

\begin{center}
\begin{tabular}{*{9}{@{\hspace{12pt}}r}}
\multicolumn{9}{c}{Table 8i. The $F^4\Delta$ $v$ = 3 Term Values, Continued}\\
{\ }\\
$J$ & \multicolumn{2}{c}{7/2} & \multicolumn{2}{c}{5/2} &
 \multicolumn{2}{c}{3/2} & \multicolumn{2}{c}{1/2}\\
    &\multicolumn{1}{c}{a} &\multicolumn{1}{c}{b}
    &\multicolumn{1}{c}{a} &\multicolumn{1}{c}{b}
    &\multicolumn{1}{c}{a} &\multicolumn{1}{c}{b}
    &\multicolumn{1}{c}{a} &\multicolumn{1}{c}{b}\\
    &\multicolumn{2}{c}{$\overbrace{\mbox{\hspace{80pt}}}$}
    &\multicolumn{2}{c}{$\overbrace{\mbox{\hspace{80pt}}}$}
    &\multicolumn{2}{c}{$\overbrace{\mbox{\hspace{80pt}}}$}
    &\multicolumn{2}{c}{$\overbrace{\mbox{\hspace{80pt}}}$}\\
36.5&  20505.96&  20392.56&  20943.16&  20826.12&  21371.23&  21250.08&  21797.99&  21672.40\\
37.5&  20838.85&  20720.44&  21280.12&  21158.08&  21713.01&  21586.90&  22145.05&  22014.57\\
38.5&  21177.98&  21054.61&  21623.21&  21496.26&  22060.76&  21929.79&  22497.87&  22362.63\\
\end{tabular}
\end{center}

\newpage

\begin{center}
\begin{tabular}{*{9}{@{\hspace{12pt}}r}}
\multicolumn{9}{c}{Table 8j. The $F^4\Delta$ $v$ = 4 Term Values}\\
{\ }\\
$J$ & \multicolumn{2}{c}{7/2} & \multicolumn{2}{c}{5/2} &
 \multicolumn{2}{c}{3/2} & \multicolumn{2}{c}{1/2}\\
    &\multicolumn{1}{c}{a} &\multicolumn{1}{c}{b}
    &\multicolumn{1}{c}{a} &\multicolumn{1}{c}{b}
    &\multicolumn{1}{c}{a} &\multicolumn{1}{c}{b}
    &\multicolumn{1}{c}{a} &\multicolumn{1}{c}{b}\\
    &\multicolumn{2}{c}{$\overbrace{\mbox{\hspace{80pt}}}$}
    &\multicolumn{2}{c}{$\overbrace{\mbox{\hspace{80pt}}}$}
    &\multicolumn{2}{c}{$\overbrace{\mbox{\hspace{80pt}}}$}
    &\multicolumn{2}{c}{$\overbrace{\mbox{\hspace{80pt}}}$}\\
 0.5&          &          &          &          &          &          &  16022.39&  16030.80\\
 1.5&          &          &          &          &  15823.88&  15831.90&  16038.60&  16046.50\\
 2.5&          &          &  15578.48&  15586.40&  15849.21&  15856.48&  16065.61&  16072.65\\
\end{tabular}
\end{center}

\begin{center}
\begin{tabular}{*{9}{@{\hspace{12pt}}r}}
\multicolumn{9}{c}{Table 8j. The $F^4\Delta$ $v$ = 4 Term Values, Continued}\\
{\ }\\
$J$ & \multicolumn{2}{c}{7/2} & \multicolumn{2}{c}{5/2} &
 \multicolumn{2}{c}{3/2} & \multicolumn{2}{c}{1/2}\\
    &\multicolumn{1}{c}{a} &\multicolumn{1}{c}{b}
    &\multicolumn{1}{c}{a} &\multicolumn{1}{c}{b}
    &\multicolumn{1}{c}{a} &\multicolumn{1}{c}{b}
    &\multicolumn{1}{c}{a} &\multicolumn{1}{c}{b}\\
    &\multicolumn{2}{c}{$\overbrace{\mbox{\hspace{80pt}}}$}
    &\multicolumn{2}{c}{$\overbrace{\mbox{\hspace{80pt}}}$}
    &\multicolumn{2}{c}{$\overbrace{\mbox{\hspace{80pt}}}$}
    &\multicolumn{2}{c}{$\overbrace{\mbox{\hspace{80pt}}}$}\\
36.5&  21486.45&  21338.48&  21916.86&  21764.08&  22337.23&  22178.95&  22755.61&  22591.33\\
37.5&  21809.96&  21655.24&  22244.42&  22084.88&  22669.63&  22504.61&  23093.36&  22922.39\\
38.5&  22139.76&  21978.35&  22578.20&  22411.96&  23008.11&  22836.43&  23437.02&  23259.45\\
\end{tabular}
\end{center}

\newpage
\begin{center}
\begin{tabular}{lr}
\multicolumn{2}{c}{Table 9a. FeH Energy Levels for Thermochemistry}\\
\\
\multicolumn{2}{c}{ }\\
{State} & {Term value / cm$^{-1}$} \\
\\
X$^4\Delta_{7/2}$ & 0 \\
X$_{5/2}$ & 191 \\
X$_{3/2}$ & 426 \\
X$_{1/2}$ & 695 \\
A$^4\Pi_{5/2}$ & 910 \\
A$_{3/2}$ & 1068 \\
A$_{1/2}$ & 1228 \\
A$_{-1/2}$ & 1388 \\
a$^6\Delta_{9/2}$ & 1766 \\
a$_{7/2}$ & 1981 \\
a$_{5/2}$ & 2177 \\
a$_{3/2}$ & 2356 \\
a$_{1/2}$ & 2518 \\
a$_{-1/2}$ & 2656 \\
C$^4\Phi_{9/2}$ & 3174 \\
C$_{7/2}$ & 3375 \\
C$_{5/2}$ & 3575 \\
C$_{3/2}$ & 3735 \\
b$^6\Pi_{7/2}$ & 3889 \\
b$_{5/2}$ & 3939 \\
b$_{3/2}$ & 3989 \\
b$_{1/2}$ & 4039 \\
b$_{-1/2}$ & 4089 \\
b$_{-3/2}$ & 4139 \\
c$^6\Sigma^+$ & 4750 \\
B$^4\Sigma^-$ & 5089 \\
D$^4\Sigma^+$ & 7158 \\
E$^4\Pi$ & 7300 \\
F$^4\Delta$ & 9995 \\
\end{tabular}
\end{center}

\newpage
\begin{center}
\begin{tabular}{lllrll}
\multicolumn{6}{c}{Table 9b. FeH Constants for Thermochemistry in cm$^{-1}$} \\
\\
State &\multicolumn{1}{c}{B$_e$} & \multicolumn{1}{c}{$\alpha_e$} &
\multicolumn{1}{c}{D$_e$} & \multicolumn{1}{c}{$\omega_e$} &
\multicolumn{1}{c}{$\omega_e x_e$} \\
\\
X$^4\Delta$ & 6.5906 & 0.2116 & $8.5 \times 10^{-5}$ & 1826.86 & 31.96 \\
A$^4\Pi$ & 6.86 & 0.22 & $1 \times 10^{-4}$ & 1875 & 34 \\
a$^6\Delta$ & 6.00 & 0.2 & $1.4 \times 10^{-4}$ & 1680 & 30 \\
C$^4\Phi$ & 6.6 & 0.2 & $1 \times 10^{-4}$ & 1680 & 30 \\
b$^6\Pi$ & 5.9 & 0.2 & $1 \times 10^{-4}$ & 1600 & 30 \\
c$^6\Sigma^+$ & 6.7 & 0.2 & $1 \times 10^{-4}$ & 1600 & 30 \\
B$^4\Sigma^-$ & 6.9 & 0.2 & $1 \times 10^{-4}$ & 1875 & 34 \\
D$^4\Sigma^+$ & 6.5 & 0.2 & $1 \times 10^{-4}$ & 1800 & 30 \\
E$^4\Pi$ & 6.86 & 0.22 & $1 \times 10^{-4}$ & 1875 & 34 \\
F$^4\Delta$ & 5.936 & 0.19 & $1 \times 10^{-4}$ & 1498 & 37 \\
\end{tabular}
\end{center}

\newpage
\begin{center}
\begin{tabular}{rll}
\multicolumn{3}{c}{Table 9c. Partition function of FeH by level
summation and analytic evaluation (see text).} \\
\\
\multicolumn{3}{c}{ } \\
{T /K}   & {Summation} & {Analytic} \\
\\
1000  &  6.97184E+02  &  8.89514E+02 \\
1100  &  8.03851E+02  &  1.08709E+03 \\
1200  &  9.17504E+02  &  1.31250E+03 \\
1300  &  1.03828E+03  &  1.56785E+03 \\
1400  &  1.16627E+03  &  1.85520E+03 \\
1500  &  1.30158E+03  &  2.17658E+03 \\
1600  &  1.44427E+03  &  2.53392E+03 \\
1700  &  1.59438E+03  &  2.92912E+03 \\
1800  &  1.75196E+03  &  3.36400E+03 \\
1900  &  1.91703E+03  &  3.84032E+03 \\
2000  &  2.08960E+03  &  4.35980E+03 \\
2100  &  2.26969E+03  &  4.92408E+03 \\
2200  &  2.45731E+03  &  5.53478E+03 \\
2300  &  2.65244E+03  &  6.19343E+03 \\
2400  &  2.85509E+03  &  6.90156E+03 \\
2500  &  3.06524E+03  &  7.66062E+03 \\
2600  &  3.28289E+03  &  8.47207E+03 \\
2700  &  3.50803E+03  &  9.33728E+03 \\
2800  &  3.74062E+03  &  1.02576E+04 \\
2900  &  3.98065E+03  &  1.12344E+04 \\
3000  &  4.22809E+03  &  1.22691E+04 \\
3100  &  4.48291E+03  &  1.33627E+04 \\
3200  &  4.74507E+03  &  1.45167E+04 \\
3300  &  5.01453E+03  &  1.57323E+04 \\
3400  &  5.29123E+03  &  1.70107E+04 \\
3500  &  5.57513E+03  &  1.83531E+04 \\
\end{tabular}
\end{center}

\newpage

\end{document}